\documentclass{article}

\usepackage{amsmath}
\usepackage{PRIMEarxiv}
\usepackage{mathrsfs}
\usepackage{listings}
\usepackage{longtable}
\usepackage{tabularx}
\usepackage{forloop}
\usepackage{etoolbox}
\usepackage{multirow}
\usepackage{subcaption}
\usepackage{hyperref} 
\usepackage{float}
\usepackage{calc}
\usepackage{wrapfig,lipsum,booktabs} % To produce the empty space on the left
\usepackage{xtab} % To produce the empty space on the left
\usepackage[dvipsnames]{xcolor}
\usepackage{tcolorbox}
\usepackage{ragged2e}
\usepackage{verbatim}
\usepackage{pdflscape}
\usepackage{changepage}
\usepackage{cuted}

\usepackage[justification=centering]{caption}
\usepackage{cuted}
\usepackage{makecell}
\usepackage{adjustbox}
\usepackage{multicol}

\usepackage[utf8]{inputenc} % allow utf-8 input
\usepackage[T1]{fontenc}    % use 8-bit T1 fonts
\usepackage{hyperref}       % hyperlinks
\usepackage{url}            % simple URL typesetting
\usepackage{booktabs}       % professional-quality tables
\usepackage{amsfonts}       % blackboard math symbols
\usepackage{nicefrac}       % compact symbols for 1/2, etc.
\usepackage{microtype}      % microtypography
\usepackage{lipsum}
\usepackage{fancyhdr}       % header
\usepackage{graphicx}       % graphics
\graphicspath{{media/}}     % organize your images and other figures under media/ folder

\usepackage[justification=centering]{caption}
\usepackage{graphicx}%
\usepackage{rotating}
\usepackage{multirow}%
\usepackage{amsmath,amssymb,amsfonts}%
\usepackage{amsthm}%
\usepackage{mathrsfs}%
\usepackage[title]{appendix}%
\usepackage{xcolor}%
\usepackage{textcomp}%
\usepackage{manyfoot}%
\usepackage{booktabs}%
\usepackage{algorithm}%
\usepackage{algorithmicx}%
\usepackage{algpseudocode}%
\usepackage{listings}%
\usepackage{hyperref}  % For handling URLs
\usepackage{url}
\usepackage{cuted}
\usepackage{longtable}
\usepackage{makecell}
\usepackage{adjustbox}
\usepackage{multicol}
\usepackage{tabularx}
\usepackage{array} 
\usepackage{natbib}
\usepackage{tikz}
\usetikzlibrary{arrows.meta, decorations.pathreplacing}

\title{Adaptive Alpha Weighting with PPO: Enhancing Prompt-Based LLM-Generated Alphas in Quant Trading}
\author{
  Qizhao Chen \\
  Graduate School of Information Science \\
  University of Hyogo \\
  Kobe, Japan\\
  \texttt{af24o008@guh.u-hyogo.ac.jp} \\
  \AND
  Hiroaki Kawashima \\
  Graduate School of Information Science \\
  University of Hyogo \\
  Kobe, Japan\\
  \texttt{kawashima@gsis.u-hyogo.ac.jp} \\
}

\begin{document}
\maketitle

\begin{abstract}
This paper introduces a reinforcement learning framework that employs Proximal Policy Optimization (PPO) to dynamically optimize the weights of multiple large language model (LLM)-generated formulaic alphas for stock trading strategies. Formulaic alphas are mathematically defined trading signals derived from price, volume, sentiment, and other data. Although recent studies have shown that LLMs can generate diverse and effective alphas, a critical challenge lies in how to adaptively integrate them under varying market conditions. To address this gap, we leverage a DeepSeek model to generate fifty alphas for ten stocks, and then use PPO to adjust their weights in real time. Experimental results indicate that the PPO-optimized strategy does not consistently deliver the highest cumulative returns across all stocks, but it achieves comparatively higher Sharpe ratios and smaller maximum drawdowns in most cases. When compared with baseline strategies, including equal-weighted, buy-and-hold, random entry/exit, and momentum approaches, PPO demonstrates more stable risk-adjusted performance. The findings highlight the importance of reinforcement learning in the allocation of alpha weights and show the potential of combining LLM-generated signals with adaptive optimization for robust financial forecasting and trading.
\end{abstract}

% keywords can be removed

\section{Introduction}\label{sec1}

Over the past few decades, financial markets have evolved significantly with the integration of advanced technologies, especially in quantitative finance. Traditionally, stock trading strategies have relied on financial indicators such as moving averages, volatility measures, and momentum indicators to predict market behavior and generate profits~\cite{9580172, MOSTAFAVI2025100631}. These methods have been successful to some extent, but often lack the flexibility and adaptability required to respond to changing market conditions. In addition, relying solely on manually chosen indicators may restrict the discovery of new patterns within the extensive and constantly changing financial data landscape.

Furthermore, trading signals often lose their effectiveness over time due to changes in market conditions and investor behavior~\cite{8279188}. This phenomenon is called Alpha Decay in quantitative finance. As a result, traders and investors must continuously search for new signals or develop new features that remain closely related to stock price movements. Staying adaptive and proactive is essential to maintain an edge in the dynamic and highly competitive financial markets.

To deal with rapidly changing and complex financial markets, advanced models and techniques have been developed to manage the increasing volume and diversity of data. One of the most promising advances in recent years is the use of machine learning, particularly deep learning models, in the development of trading strategies~\cite{MINTARYA202396}. These models can process large datasets and uncover complex patterns in market behavior that traditional methods might miss. However, despite their impressive capabilities, machine learning models still face challenges in areas such as interpretability~\cite{Sejnowski_2020}, flexibility, and the need for extensive data labeling~\cite{10114634,kou2024automatestrategyfindingllm}.

A new development in this field involves the application of large language models (LLMs), such as OpenAI's GPT models, for financial analysis~\cite{Chen2025PromptLLM}. LLMs, which have been trained on large amounts of text data, can process financial news, reports, and historical price information to generate insights and predictions~\cite{10825946}. These models can generate a variety of outputs, including summarizing financial news and creating algorithmic strategies for trading. In particular, prompt-based LLMs, which take in specific prompts and generate relevant outputs based on those inputs, provide a powerful tool for automating and enhancing the alpha generation process in stock trading~\cite{wang2024gptsignalgenerativeaisemiautomated}. However, most of the time, other researchers only use LLMs to generate formulaic alphas and examine their correlations with stock return. They do not analyze how the LLM-generated alphas can be used in dynamic quant trading, which will be further investigated in this paper.

In our previous study~\cite{chen2025sentimentawarestockpriceprediction}, LLM-generated formulaic alphas have been demonstrated to have strong predictive power in stock price prediction. In this study, we extend this line of research by shifting the focus from alpha generation to alpha integration and optimization. Instead of relying solely on the predictive strength of individual alphas, we investigate how reinforcement learning, specifically Proximal Policy Optimization, can be employed to dynamically allocate weights across multiple LLM-generated alphas. This approach enables the trading strategy to adapt to evolving market conditions, enhance robustness against noisy signals, and capture diverse market dynamics.

First, we use a prompt-based LLM, deepseek-r1-distill-llama-70b model, to generate formulaic alphas for ten major companies: Toyota, Apple, HSBC, Pepsi, Tencent, Airbus, Exxon Mobil, Petrobras, Netfilx, and InfuSystem. Alpha, in strict financial definition, refers to the excess return that a trading strategy can achieve above a benchmark index, such as the S\&P 500 or Nikkei 225~\cite{kakushadze2016101formulaicalphas}. In quantitative finance and machine learning, alpha factors often refer to signals or features that are predictive of future stock returns. Formulaic alpha involves mathematically defined trading signals or strategies aimed at predicting future asset returns. These strategies are usually expressed as explicit formulas that combine various types of data, including market, fundamental, or alternative inputs such as price, volume, volatility, and sentiment indicators. Traditionally, these formulas often originate from quantitative research. Equation~\ref{eq:mean_reversion_alpha} is an example of formulaic alpha. When a stock price is substantially higher than its recent average, Equation~\ref{eq:mean_reversion_alpha} yields a negative value, indicating that the stock might be overbought and likely to decline. However, if the price is below the average, the alpha becomes positive, suggesting a potential buying opportunity due to mean reversion.

\begin{equation}
\alpha_t = -(\text{Close}_t - \text{SMA}_{10,t})
\label{eq:mean_reversion_alpha}
\end{equation}

\begin{itemize}
  \item $\text{Close}_t$: The closing price of the stock at time $t$.
  \item $\text{SMA}_{10,t}$: The 10-day Simple Moving Average of the stock's closing price at time $t$.
\end{itemize}

Once formulaic alphas are generated, this study applies proximal policy optimization (PPO) to optimize the weights of each alpha. The reason for combining LLM-generated alphas with reinforcement learning is that recent studies often treat each LLM-generated alpha independently or combine them using static or manually designed weighting schemes. However, financial markets are dynamic and often non-stationary, meaning that the effectiveness of individual alphas can vary significantly over time. To address this, we propose combining LLM-generated alphas with reinforcement learning to dynamically learn optimal alpha weights based on changing market conditions. This integration allows the model to treat alpha weighting as a sequential decision-making task, where the agent continuously adjusts its strategy in response to observed market feedback. Compared to traditional methods such as regression-based weighting or simple averaging, our approach enables the system to adaptively prioritize more informative or better performing alphas in real time, thus improving overall trading performance and robustness.

PPO, a model-free algorithm proposed by~\cite{schulman2017proximalpolicyoptimizationalgorithms}, is particularly well-suited for continuous action spaces, making it ideal for determining the optimal weight distribution of different alphas in a portfolio. Unlike off-policy methods such as Deep Deterministic Policy Gradient (DDPG) and Soft Actor-Critic (SAC), which are often sensitive to hyperparameter choices and can struggle in non-stationary environments, PPO employs a clipped surrogate objective function to regulate policy updates. This design effectively limits the magnitude of each update, acting as a trust-region–like constraint that helps avoid abrupt changes in the trading strategy, which are particularly problematic under the low signal-to-noise conditions of financial markets~\cite{10.5555/3491440.3492067}. Moreover, by adopting an on-policy learning framework, PPO promotes more stable learning dynamics~\cite{ZHANG2022750} and a balanced exploration–exploitation behavior, reducing the tendency to adapt excessively to short-term market noise. For example, Rio et al.~\cite{10703056} show that PPO has a more stable training process compared to Asynchronous Advantage Actor-Critic (A3C). Mammadzada~\cite{Mammadzada2025PPO} finds that PPO has more stable and consistent performance compared to SAC. Consequently, by training the PPO agent, this study seeks to identify a robust combination of alphas that maximizes risk-adjusted returns while ensuring training stability across varying market regimes.

To evaluate the performance of our strategy, we compare its returns with those of several baseline approaches, including an equal-weighted portfolio, a buy-and-hold strategy, a random entry/exit strategy, and a momentum strategy. In addition to cumulative returns, we also analyze the Sharpe ratio and maximum drawdown to provide a more comprehensive view of performance. The Sharpe ratio reflects the risk-adjusted return of the strategy, helping to assess whether higher returns are achieved efficiently. Maximum drawdown captures the largest peak-to-trough decline, indicating potential risk exposure during unfavorable market conditions. By evaluating these metrics together, we aim to determine not only whether LLM-generated alphas can outperform traditional market indices but also whether they can do so with more favorable risk-return characteristics.

The primary goal of this research is to demonstrate the potential of combining LLMs for alpha generation with reinforcement learning techniques such as PPO to create more adaptive and dynamic trading strategies. By exploring how machine learning models can work together with financial theory, this study offers insights into how technology can transform financial decision-making and trading practices. 

The main contributions of this paper are as follows.

\begin{enumerate}

    \item This study introduces a reinforcement learning framework using PPO to dynamically optimize the weights of multiple LLM-generated alphas, adapting to changing market conditions.

    \item Through an ablation study comparing human-crafted and LLM-generated alphas, this study demonstrates the superior performance of LLM-generated alphas.

    \item Empirical analysis is conducted across multiple alpha selection strategies, including random, low-correlation, and high-contribution methods, demonstrating the consistent performance of the proposed framework across different stocks. This provides practical insights into how alpha quality and selection criteria influence portfolio performance.

\end{enumerate}

The remainder of this paper is structured as follows. Firstly, related work is listed (Section 2). The methodology is then described (Section 3). In Section 4, the experimental results are presented. Further discussion is provided in Section 5. Finally, Section 6 concludes the paper.

\section{Related Work}\label{sec2}

The concept of alpha generation in finance refers to the ability of a trading strategy to outperform the market or a benchmark, such as stock indices. Traditional alpha generation methods are based on fundamental and technical analysis, which identified patterns in historical stock prices, trading volumes, and other economic factors. However, as financial markets become more complex, quantitative approaches leveraging machine learning, and more recently deep learning, have been increasingly used to develop formulaic alphas, which are mathematical formulas or models designed to predict market movements and generate excess returns.

Early efforts in formulaic alpha generation rely on statistical and econometric models. For example, Fama and French~\cite{FAMA19933} propose factor models, which aim to explain asset returns using factors such as market risk, size, value, and momentum. These models have been foundational in understanding systematic risk factors, but they have limitations in capturing complex, non-linear relationships in financial data. More recent advancements have focused on enhancing traditional alpha models by incorporating machine learning algorithms.

Machine learning approaches have gained significant attention in the domain of alpha generation. One notable development is the use of supervised learning models to predict stock prices and returns. Researchers have applied various machine learning techniques, including decision trees~\cite{8250694}, support vector machines (SVMs)~\cite{6703096,9361804}, and random forests~\cite{9987903}, to identify patterns in historical data and generate alpha signals. However, these models are often limited by their reliance on structured data and the need for manual feature engineering, which can be time-consuming and prone to biases~\cite{chen2025llmstock}.

A more recent shift in alpha generation research has involved the application of deep learning techniques, particularly convolutional neural networks (CNNs)~\cite{9317207,chen2025image} and recurrent neural networks (RNNs)~\cite{10392023}, to stock market prediction. These models are capable of automatically learning complex and non-linear patterns from raw market data. Recently, some researchers have proposed some hybrid models to further improve prediction performance. For example, Zhang et al.~\cite{math11091985} propose a CNN-BiLSTM-Attention-based model to improve the performance of stock price forecasting. The proposed approach can outperform models such as LSTM. Lu et al.~\cite{https://doi.org/10.1155/2020/6622927} propose a CNN-LSTM to predict the stock price.

Recently, transformer-based models have become popular candidates in stock price forecasting~\cite{chen2025anomaly}. Some transformer models designed for time series forecasting include Dual Transformer~\cite{chen2025sentiment}, Informer~\cite{zhou2021informerefficienttransformerlong}, Autoformer~\cite{wu2022autoformerdecompositiontransformersautocorrelation} and TimeXer~\cite{wang2024timexerempoweringtransformerstime}. For example, Li et al.~\cite{li2023mastermarketguidedstocktransformer} proposed a market-guided transformer to predict the stock price. This model uses momentary and cross-time stock correlation as well as market information for automatic feature selection.

In addition to supervised learning and deep learning, natural language processing (NLP) techniques have been used to extract information from unstructured data sources such as financial news articles, earnings reports, and social media~\cite{chen2025frameworkmeasuringnewstopics}. NLP models, especially generative LLMs such as LLaMA, Mistral, and GPT, have shown promise in sentiment analysis and in predicting stock movements based on textual data~\cite{10825946, yang2023fingptopensourcefinanciallarge,xie2023pixiulargelanguagemodel}. These models analyze the sentiment of news and social media content and generate alpha signals based on the prevailing market sentiment. This approach has become more prominent due to the increasing availability of financial news and the growing importance of market sentiment in stock price movements.

Prompt-based LLMs are another emerging area of interest in formulaic alpha generation. These models generate specific financial alphas based on customized input prompts. For example, researchers have explored the use of prompt-based LLMs to generate alphas that incorporate technical indicators, such as moving averages, RSI, and MACD, along with sentiment data from news articles or earnings calls~\cite{kou2025automatestrategyfindingllm}. For example, Chen and Kawashima~\cite{chen2025sentimentawarestockpriceprediction} utilize an LLM to generate formulaic alphas and provide evidence of their effectiveness in predicting stock price movements. Wang et al.~\cite{wang2024gptsignalgenerativeaisemiautomated} use the prompt-based GPT-4 model to generate formulaic alphas. By tailoring the prompts to include specific financial metrics, these LLMs can generate alphas that are directly aligned with the desired prediction task. This approach differs from traditional machine learning techniques because it combines domain-specific knowledge (in the form of technical indicators) with the generative capabilities of LLMs to create actionable trading strategies. However, their research only generates alphas but does not address how to adaptively combine them or assess their performance over time.

Reinforcement learning (RL) has gained increasing attention in financial forecasting and trading. RL models are designed to learn optimal trading strategies by interacting with the market environment through trial and error. By continuously updating their policies based on observed rewards, RL algorithms can adapt to changing market dynamics and improve decision making over time~\cite{Chen2025}. Various RL methods, such as Deep Q-Networks (DQN), Proximal Policy Optimization (PPO), and Soft Actor-Critic (SAC), have been successfully applied to tasks like portfolio management, asset allocation, and algorithmic trading~\cite{huang2025deepreinforcementlearningframework,ndikum2024advancinginvestmentfrontiersindustrygrade,10627674,ijfs11010010}. These approaches allow trading agents to make sequential decisions, balancing the trade-off between risk and return in complex and uncertain markets. For example, Lee and Moon~\cite{10285085} propose a transformer-based actor–critic reinforcement learning model that uses past stock price information to guide trading decisions and shows improved risk-adjusted returns across multiple stock market datasets compared with existing methods. Orra et al.~\cite{orra2025reward} introduce an expert-guided reward design for deep reinforcement learning in stock trading, where technical indicator signals provide feedback to stabilize learning and improve trading performance across multiple global stock markets. Kong and So~\cite{app13010633} extend ensemble-based reinforcement learning for stock trading by combining multiple actor–critic methods into a larger ensemble and applying it across several global stock markets, showing more stable returns in empirical tests. Guevara~\cite{guevara2025actorcritic} proposes an actor–critic reinforcement learning trading model with an adaptive data window, showing reduced losses and faster learning across crude oil, gold, and Euro markets. Sattar et al.~\cite{10904473} propose a reinforcement learning trading framework that combines price data, sentiment indicators, and drawdown-based rewards, showing improved risk control and returns compared with baseline market strategies.

Although reinforcement learning has been extensively applied in finance, the integration of LLMs and RL for optimizing multiple formulaic alphas is still a relatively novel area of research. Few studies have directly addressed the use of LLMs for alpha generation in combination with RL algorithms for weight optimization. We will examine the effectiveness of this framework in this study.

\section{Methodology}\label{sec3}

\begin{figure*}[!htbp]
    \centering
    \includegraphics[width=1\textwidth]{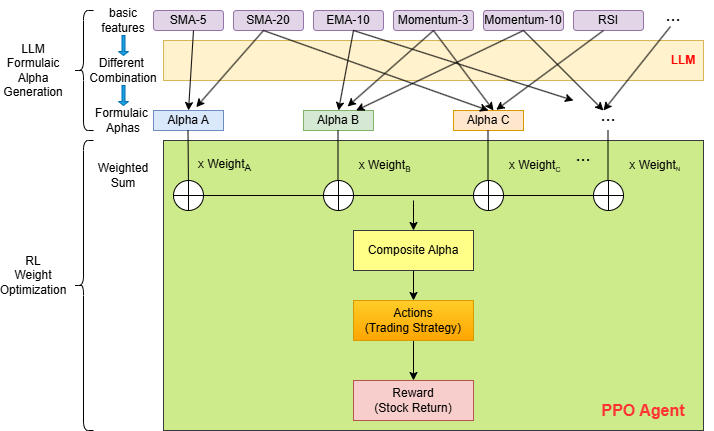}  % Adjust the width as needed
    \caption{Whole Picture of the Proposed Method}
    \label{fig: whole picture}
\end{figure*}

In this study, we focus on generating formulaic alphas for predicting stock returns using a combination of prompt-based LLMs and reinforcement learning to optimize the weights of each alpha. The goal is to develop a framework that can produce alpha signals for multiple companies and optimize these signals to improve trading performance. Figure~\ref{fig: whole picture} illustrates the setup of this research.

\begin{figure*}[!htbp]
    \centering
    \includegraphics[width=1\textwidth]{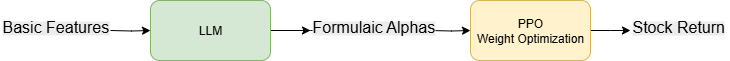}  % Adjust the width as needed
    \caption{Flow Chart of the Proposed Method}
    \label{fig: flow chart}
\end{figure*}

\subsection{Data}
The data used for this research come from two primary channels: stock data and financial news sentiment data. 

Since multiple related studies~\cite{VIJH2020599,electronics9091384,10195852,10.1145/3529836.3529857} in stock forecasting or trading typically focus on one to ten stocks, this study selects ten stocks to balance representativeness and computational cost. The selected stocks—Toyota, Apple, HSBC, Pepsi, Tencent, Airbus, Exxon Mobil, Netflix, Petrobras, and InfuSystem—come from different regions, including the U.S., Europe, China, Japan, and Brazil, and cover a broad range of industries such as automotive, technology, finance, energy, aerospace, healthcare services, and consumer goods. This diversity helps evaluate whether the proposed method works across different market conditions rather than for a single sector. Historical stock price data from 2016/02/16 to 2024/05/08 are obtained using the yfinance Python library. This library provides an efficient way to access daily stock prices, including opening, closing, high, low, and volume data for each company. 

In addition, stock data is used to calculate key technical indicators, such as moving averages, momentum, and RSI, which are incorporated into the alpha generation process. The Python library pandas-ta is used to calculate all the technical indicators. In finance, technical indicators, calculated from historical stock price and volume data, are often used by traders and investors to predict the stock trend and generate trading signals. Table~\ref{tab:tech_indicators} shows the technical indicators used in this study.

The \textbf{Simple Moving Average (SMA)} calculates the average price of a security over a specified period, smoothing out short-term fluctuations to reveal overall trends. In contrast, the \textbf{Exponential Moving Average (EMA)} places greater emphasis on recent prices, making it more responsive to new information compared to the SMA. \textbf{Momentum} measures the speed of price changes by comparing the current price to a previous price from a set number of periods ago, helping traders identify potential trend changes. The \textbf{Relative Strength Index (RSI)} is a momentum oscillator that assesses the speed and change of price movements, commonly used to detect overbought or oversold conditions. \textbf{MACD (Moving Average Convergence Divergence)} tracks the relationship between two EMAs, helping traders understand changes in the strength, direction, and momentum of a trend. Bollinger Bands are plotted two standard deviations away from a moving average, offering a visual representation of price volatility and potential reversal points. \textbf{On-Balance Volume (OBV)} combines price movement with volume flow to indicate how volume affects the direction of the price.

\begin{table}[!htbp]
\centering
\caption{Technical Indicators Used for Alpha Generation}
\label{tab:tech_indicators}
\begin{tabular}{|c|}
\hline
\textbf{Indicator Name} \\
\hline
Simple Moving Average (5-day) \\
Simple Moving Average (20-day) \\
Exponential Moving Average (10-day) \\
Momentum (3-day) \\
Momentum (10-day) \\
Relative Strength Index (14-day) \\
Moving Average Convergence Divergence (MACD) \\
MACD Signal Line \\
Bollinger Band Upper \\
Bollinger Band Lower \\
On-Balance Volume (OBV) \\
\hline
\end{tabular}
\end{table}

Daily news articles related to target companies are downloaded from websites such as Yahoo News using the Financial News Feed and Stock News Sentiment Data API~\footnote{https://eodhd.com/financial-apis/stock-market-financial-news-api}. The sentiment of these articles is computed using a natural language processing model, NLTK that analyzes the tone of the text and outputs a polarity score. The polarity score is used to quantify sentiment, ranging from $-1$ to $1$. A score close to $1$ indicates strong positive sentiment, while a score close to $-1$ reflects strong negative sentiment. A score of $0$ represents neutral sentiment. This sentiment score is then used to influence the generation of alphas, as market sentiment can significantly impact stock prices. The sentiment data are integrated into the LLM prompt, along with historical price and technical indicator data, to generate formulaic alphas for each company.

The first 80\% of the data will be used for training and the remaining 20\% of the data will be used for testing. We adopt an 80/20 split between the training and test sets to achieve a balanced trade-off between model learning and out-of-sample evaluation. Prior work by~\cite{https://doi.org/10.1002/sam.11583} suggests an optimal test proportion of approximately $1/(\sqrt{p}+1)$. In our setting, where $p=50$ alpha features are used as predictors, this guideline corresponds to a test size of about 12.3\%.

We deliberately choose a larger test split of 20\% for a practical reason. From a statistical standpoint, a larger hold-out period enables a more reliable evaluation of performance metrics such as the Sharpe Ratio and Maximum Drawdown, reducing the likelihood that the results are driven by short-term market fluctuations.

In Section~\ref{sec: walk-forward analysis}, we also conduct a walk-forward optimization analysis to further examine the robustness of the reinforcement learning framework.

\subsection{Model Setup}

\subsubsection{\textbf{Prompt-Based LLM for Alpha Generation}}
The model used to generate formulaic alphas is a prompt-based LLM, specifically deepseek-r1-distill-llama-70b. This LLM is provided by American AI company Groq\footnote{https://groq.com/} for deployment and inference. This version of DeepSeek model has been shown to generate competitive results without fine-tuning compared to other LLMs such as LLaMA and Gemma in previous study~\cite{Chen2025}. This model (deepseek-r1-distill-llama-70b) uses knowledge distillation, a process in which a smaller model (the student) learns from the output of a larger, more complex model (the teacher). In this case, the LLaMA-70B model serves as the teacher, and the student model is trained to replicate its behavior. The teacher model generates predictions or soft targets, which guide the student in learning more efficiently. Knowledge distillation allows the student model to retain much of the performance of the teacher model while being more lightweight, making it suitable for deployment in environments with limited resources~\cite{hinton2015distillingknowledgeneuralnetwork}. 

The input data fed to the prompt (\texttt{\{features\}} in Table~\ref{tab:LLM_prompt}) include historical prices, technical indicators, and sentiment scores of all companies. The input data are stored in a pandas DataFrame and then converted to JSON format. The model generates a common set of fifty distinct formulaic alphas shared across all companies. These alphas are mathematical expressions that combine various financial metrics, such as momentum, moving averages, RSI, MACD, and Bollinger Bands. The LLM is prompted with historical data and a detailed description of the task, which guide it to generate alphas that can be used to predict stock movements.

For example, one formula might combine momentum indicators with moving averages, while another might include sentiment scores to account for market news. Fifty formulas are generated for the forecasting of stock returns of all companies. To avoid look-ahead bias, only training data is used to generate formulaic alphas.

\begin{table*}[!htbp]
\centering
\caption{LLM Prompt for Formulaic Alpha Generation}
\label{tab:LLM_prompt}
\begin{tabular}{|p{0.95\textwidth}|}
\hline
\texttt{You are a quantitative trader. Generate 50 alpha formulas using the given stock features:} 
\texttt{\{features\}.} \\
\texttt{The formulas should be mathematical expressions combining these features.} \\
\texttt{Return only the formulas in Python syntax, using variables like C\_t (Close), O\_t (Open),} 
\texttt{V\_t (Volume), S\_t (Sentiment), and standard indicators (SMA, Momentum).} \\
\texttt{Example Output: alpha\_t = (C\_t - O\_t) / O\_t + 0.5 * S\_t} \\
\hline
\end{tabular}
\end{table*}

\subsubsection{\textbf{PPO for Alpha Weight Optimization}}

The alpha weight optimization process employs PPO, a reinforcement learning algorithm particularly suited for continuous action spaces in financial applications. This choice is motivated by PPO's three key properties: (1) its clipped objective function prevents destructively large policy updates, (2) its sample efficiency aligns with the limited availability of high-quality financial data, and (3) its ability to handle the non-stationary dynamics inherent in market environments.

The trading environment constitutes a partially observable Markov decision process (POMDP) where the state space $s_t \in \mathcal{S}$ at time $t$ integrates four information categories:

\begin{equation}
s_t = \{\text{OHLCV}_t,\, p_{t-1},\, \text{regime}_t,\, \sigma_t\}
\label{eq:state_representation}
\end{equation}

\noindent Here, $\text{OHLCV}_t$ (Open, High, Low, Close, Volume) represents the raw price and volume data, $p_{t-1} \in [-1,1]$ denotes the previous position, $\text{regime}_t \in \{0,1\}$ indicates the bull/bear market classification based on the 20-day and 100-day moving average crossover, and $\sigma_t$ captures the 63-day rolling daily volatility annualized via $\sigma_t^{\text{annual}} = \sigma_t^{\text{63d}} \times \sqrt{252}$. The action space $\mathcal{A}$ consists of 50-dimensional weight vectors $\mathbf{w}_t \in \mathbb{R}^{50}$ corresponding to each alpha signal. These weights undergo normalization through an $L_1$-constrained transformation:

\begin{equation}
\tilde{\mathbf{w}}_t = \operatorname{clip}(\mathbf{w}_t, -1, 1)
\label{eq:weight_tilted}
\end{equation}

\begin{equation}
\mathbf{w}_t^{\text{norm}} =
\frac{\tilde{\mathbf{w}}_t}{\|\tilde{\mathbf{w}}_t\|_1 + \epsilon},
\qquad \epsilon = 10^{-8}
\label{eq:weight_normalized}
\end{equation}

\noindent Clipping ensures that each element of the weight vector $\mathbf{w}_t$ lie in the interval $[-1,1]$, which prevents extreme weights and enforces a bounded action space at the individual weight level. After clipping, the weight vector is divided by its $L_1$-norm $\|\tilde{\mathbf{w}}_t\|_1$, which ensures that the sum of the absolute values of the normalized weights is approximately 1. This is like requiring the agent to allocate weights across signals such that the total exposure is normalized to 1. This normalization preserves the interpretability of long/short signals while maintaining numerical stability by adding the $\epsilon$ to avoid division by zero. 

The reward function $r_t$ combines two critical components:

\begin{equation}
r_t = \underbrace{p_t \cdot R_t^{\text{future}}}_{\text{Position P\&L}} - \underbrace{\lambda |p_t - p_{t-1}|}_{\text{Transaction Cost}} 
\label{eq:reward_function}
\end{equation}

\noindent where $R_t^{\text{future}}$ denotes the realized return of the asset over the next period, measured as the relative price change from $t$ to $t+1$. $\lambda=0.1\%$ quantifies the transaction cost. The position $p_t$ derives from a multi-stage calculation

\begin{equation}
    \alpha^{\text{composite}}_t = \sum_{i=1}^{50} w_t[i] \cdot \alpha_{i,t}
    \label{eq:composite alpha}
\end{equation}

\begin{equation}
p_t = v_t \cdot \begin{cases}
\min(1, 2(\alpha_t^{\text{composite}} - \tau_t^{\text{upper}})) & \text{if } \alpha_t^{\text{composite}} > \tau_t^{\text{upper}} \\
\max(-1, 2(\alpha_t^{\text{composite}} - \tau_t^{\text{lower}})) & \text{if } \alpha_t^{\text{composite}} < \tau_t^{\text{lower}} \\
0 & \text{otherwise}
\end{cases}
\label{eq:position_scaling}
\end{equation}

\noindent with $v_t = \min(2, \sigma_{\text{target}}/\sigma_t^{\text{annual}})$ implementing volatility scaling ($\sigma_{\text{target}}=0.15$) and $\tau_t$ representing quintile thresholds. $\alpha_t^{\text{composite}}$ is the composite alpha, computed as a weighted sum of the individual alpha signals at time $t$ (Equation~\ref{eq:composite alpha}). Each individual alpha is standardized with the \texttt{StandardScaler} prior to computing the composite alpha.

The volatility scaling factor $v_t$ dynamically adjusts position sizes to maintain stable risk exposure. When realized volatility ($\sigma_t^{\text{annual}}$) exceeds the target 15\%, positions are scaled down to prevent overexposure during turbulent markets. Conversely, positions can scale up in low-volatility regimes to capitalize on opportunities without exceeding the risk budget.

The rolling quintile thresholds $\tau_t^{\text{upper}} = Q_{0.75}(\text{Close}_{t-126:t})$ and $\tau_t^{\text{lower}} = Q_{0.25}(\text{Close}_{t-126:t})$ serve as adaptive signal filters. Using the 75th and 25th percentiles of recent prices, these thresholds automatically adjust to market conditions. They widen when prices swing wildly (needing stronger signals before trading) and tighten when prices move steadily (capturing smaller opportunities). This prevents excessive trading during unpredictable back-and-forth price movements while still responding to real trends. The 126-day window balances long enough to filter out temporary market noise but short enough to adapt to changing conditions.

We further elaborate on the definition and role of the market regime in our framework. To incorporate regime-aware risk control, we introduce a soft penalty when the trading position is inconsistent with the prevailing market regime.
The market regime indicator $\text{regime}_t \in \{0,1\}$ is defined based on the moving-average crossover,
where $\text{regime}_t = 1$ denotes a bullish regime (20-day MA above 100-day MA) and $\text{regime}_t = 0$ indicates a bearish regime.

A regime violation occurs when the sign of the position conflicts with the regime signal, i.e.,
\begin{equation}
\mathbb{I}_t^{\text{regime}} =
\begin{cases}
1, & (\text{regime}_t = 0 \ \wedge \ p_t > 0) \ \text{or} \ (\text{regime}_t = 1 \ \wedge \ p_t < 0), \\
0, & \text{otherwise},
\end{cases}
\label{eq:regime_indicator}
\end{equation}

\noindent and the corresponding regime penalty is defined as
\begin{equation}
\mathcal{P}_t^{\text{regime}} = \lambda_{\text{reg}} \cdot |p_t| \cdot \mathbb{I}_t^{\text{regime}},
\label{eq:regime_penalty}
\end{equation}
where $\lambda_{\text{reg}}$ controls the strength of the penalty. In this study, $\lambda_{\text{reg}}$ equals to 0.05.

The PPO objective function incorporates these elements through

\begin{equation}
\mathcal{L}^{\text{CLIP}}(\theta) = \mathbb{E}_t\left[\min\left(\frac{\pi_\theta(a_t|s_t)}{\pi_{\theta_{\text{old}}}(a_t|s_t)} \hat{A}_t,\ 
\text{clip}\left(\frac{\pi_\theta(a_t|s_t)}{\pi_{\theta_{\text{old}}}(a_t|s_t)},\ 1-\epsilon,\ 1+\epsilon\right)\hat{A}_t\right)\right]
\label{eq:ppo_objective}
\end{equation}

\noindent where $\epsilon=0.2$ defines the clipping range and $\hat{A}_t$ denotes the advantage estimate computed using generalized advantage estimation (GAE). This formulation enables stable weight updates while accounting for the non-IID nature of financial time series through proper advantage discounting.

The PPO agent was trained using the default hyperparameters of Stable-Baselines3~\footnote{https://stable-baselines3.readthedocs.io/en/master/}, with learning rate $3 \times 10^{-4}$, rollout length 2048, batch size 64, 10 update epochs per rollout, discount factor 0.99, and clipping parameter 0.2.

Table~\ref{tab:algorithm} shows the whole picture of the alpha trading algorithm and Table~\ref{tab:params} illustrates the key parameters and their ranges of values.

\begin{table*}[!htbp]
\centering
\caption{Alpha Trading Algorithm with Adaptive Risk Controls}
\label{tab:algorithm}
\begin{tabular}{>{\raggedright}p{\textwidth}}
\toprule
\textbf{Algorithm: PPO-based Alpha Trading Environment} \\

\textbf{Inputs}: \\
\quad Price data $\mathcal{D} = \{OHLCV_t, R^{\text{future}}_t\}_{t=1}^T$ \\
\quad Alpha signals $\{\alpha_{i,t}\}_{i=1}^{50}$ \\
\quad Volatility target $\sigma_{\text{target}} = 0.15$ \\
\quad Transaction cost coefficient $\lambda$ \\
\quad Regime penalty coefficient $\lambda_{\text{reg}}$ \\[4pt]

\textbf{Outputs}: \\
\quad Portfolio value series $\{V_t\}_{t=1}^T$ and performance metrics $\mathcal{M}$ \\[6pt]

\textbf{Initialization}: \\
1. \textit{Feature preprocessing}: \\
\quad $\alpha_{i,t} \leftarrow \text{StandardScaler}(\alpha_{i,t}),\ \forall i \in \{1,\dots,50\}$ \\
\quad $\text{MA}_{20,t} \leftarrow \text{MA}(\text{Close}_t, 20), \quad
\text{MA}_{100,t} \leftarrow \text{MA}(\text{Close}_t, 100)$ \\
\quad $\text{regime}_t \leftarrow \mathbb{I}(\text{MA}_{20,t} > \text{MA}_{100,t})$ \\
\quad $\sigma^{\text{daily}}_t \leftarrow \text{Std}(R^{\text{future}}_{t-62:t})$ \\
\quad $\sigma^{\text{annual}}_t \leftarrow \sigma^{\text{daily}}_t \sqrt{252}$ \\
\quad $\tau^{\text{upper}}_t \leftarrow Q_{0.75}(\text{Close}_{t-126:t}), \;
\tau^{\text{lower}}_t \leftarrow Q_{0.25}(\text{Close}_{t-126:t})$ \\[4pt]

2. \textit{State initialization}: \\
\quad $V_0 \leftarrow 1.0,\; p_0 \leftarrow 0,\; \text{peak} \leftarrow 1.0$ \\[6pt]

\textbf{Trading Loop}: \\
3. \textbf{for} $t = 1$ \textbf{to} $T-1$ \textbf{do}: \\

\quad 4. \textit{Policy action}: \\
\qquad $\mathbf{w}_t \leftarrow \pi_\theta(s_t)$ \\
\qquad $\tilde{\mathbf{w}}_t \leftarrow \operatorname{clip}(\mathbf{w}_t, -1, 1)$ \\
\qquad $\mathbf{w}_t^{\text{norm}} \leftarrow
\tilde{\mathbf{w}}_t / (\|\tilde{\mathbf{w}}_t\|_1 + \epsilon)$ \\

\quad 5. \textit{Composite alpha construction}: \\
\qquad $\alpha^{\text{composite}}_t \leftarrow
\sum_{i=1}^{50} w_t^{\text{norm}}[i] \cdot \alpha_{i,t}$ \\

\quad 6. \textit{Signal-based position sizing}: \\
\qquad $p_t \leftarrow
\begin{cases}
\min(1, 2(\alpha^{\text{composite}}_t - \tau^{\text{upper}}_t)), & \alpha^{\text{composite}}_t > \tau^{\text{upper}}_t \\
\max(-1, 2(\alpha^{\text{composite}}_t - \tau^{\text{lower}}_t)), & \alpha^{\text{composite}}_t < \tau^{\text{lower}}_t \\
0, & \text{otherwise}
\end{cases}$ \\[4pt]

\qquad \textit{Volatility targeting}: \\
\qquad $v_t \leftarrow \min(2, \sigma_{\text{target}} / \sigma^{\text{annual}}_t)$ \\
\qquad $p_t \leftarrow p_t \cdot v_t$ \\[4pt]

\qquad \textit{Regime-aware risk penalty}: \\
\qquad $\mathcal{P}_t^{\text{regime}} \leftarrow
\lambda_{\text{reg}} |p_t| \cdot
\mathbb{I}(\text{sign}(p_t) \neq \text{regime}_t)$ \\

\quad 7. \textit{Reward and portfolio update}: \\
\qquad $\text{TC}_t \leftarrow \lambda |p_t - p_{t-1}|$ \\
\qquad $r_t \leftarrow p_t R^{\text{future}}_t - \text{TC}_t - \mathcal{P}_t^{\text{regime}}$ \\
\qquad $V_t \leftarrow V_{t-1} (1 + r_t)$ \\
\qquad $\text{peak} \leftarrow \max(\text{peak}, V_t)$ \\
\qquad $\text{DD}_t \leftarrow (V_t - \text{peak}) / \text{peak}$ \\

\quad 8. \textit{State transition}: \\
\qquad $s_{t+1} \leftarrow \{\text{OHLCV}_{t+1}, p_t, \text{regime}_{t+1}, \sigma^{\text{daily}}_{t+1}\}$ \\

\textbf{end for} \\[6pt]

\textbf{Evaluation}: \\
9. Compute out-of-sample metrics (Sharpe ratio, max drawdown, turnover). \\

\end{tabular}
\end{table*}

\begin{table*}[!h]
\centering
\caption{Key Parameters and Their Value Ranges}
\label{tab:params}
\begin{tabular}{lll}
\toprule
\textbf{Parameter} & \textbf{Description} & \textbf{Typical Range / Value} \\
\midrule
$\sigma_{\text{target}}$ & Annualized volatility target & 0.15 (fixed) \\
$\text{TC}_t$ & Transaction cost per trade & 0.001 \\
$\alpha_{i,t}$ & Alpha signals & Normalized to $\sim \mathcal{N}(0,1)$ \\
$\epsilon$ & Stability constant in normalization & $10^{-8}$ \\
$p_t$ & Position sizing value & $[-1, 1]$ \\
$v_t$ & Volatility scaling factor & $[0, 2]$ \\
$\tau^{\text{upper}}_t$ & Upper threshold (75\% quantile) & Time-varying \\
$\tau^{\text{lower}}_t$ & Lower threshold (25\% quantile) & Time-varying \\
$w_t[i]$ & Alpha weight from model & $[-1, 1]$, normalized \\
$\text{MA}_{20}, \text{MA}_{100}$ & Moving averages for regime filter & Computed over 20 and 100 days \\
$\sigma_{\text{daily}}$ & Daily return volatility & Rolling 63-day std. dev. \\
$\sigma_{\text{annual}}$ & Annualized volatility & $\sigma_{\text{daily}} \times \sqrt{252}$ \\
\bottomrule
\end{tabular}
\end{table*}

\subsection{Implementation Details}

\begin{tikzpicture}[node distance=2cm, font=\small]
    % Draw timeline axis
    \draw[->, thick] (0,0) -- (10,0) node[right] {Time};

    % Draw time ticks
    \foreach \x/\t in {1/t-1, 4/t, 8/t+1} {
        \draw[thick] (\x,0.1) -- (\x,-0.1) node[below] {$\t$};
    }

    % Info Set Brackets
    \draw[decorate, decoration={brace, amplitude=5pt}] (0,0.5) -- (4,0.5) 
        node[midway, above=6pt, text width=3cm, align=center] {Information Set \\ (Regime, Vol, LLM Alphas)};

    % Decision point at t
    \filldraw[blue] (4,0) circle (2pt);
    \node[blue, above right] at (4,0) {\textbf{Signal / Position $p_t$}};

    % Holding Period Arrow
    \draw[line width=1.5pt, red, -{Stealth}] (4,-0.8) -- (8,-0.8) 
        node[midway, below] {Holding Period ($p_t$)};

    % Return Realization
    \node[align=center] at (8, 1) {Return $r_{t+1}$ \\ realized at close};
    \draw[->, dashed] (8, 0.7) -- (8, 0.1);

    % Equation Mapping
    \node at (6, -1.8) {\textbf{Return Attribution:} $R_{t+1} = p_t \cdot r_{t+1}$};
\end{tikzpicture}

At each trading day $t$, the agent makes a trading decision based solely on information available up to the end of day $t$. The information set includes regime indicators, realized volatility estimates, and standardized LLM-generated alpha signals computed from historical data only. Trades are executed at the close price of day $t$, and no information beyond day $t$ is used when determining trading positions.

The agent outputs a vector of weights over the LLM-generated alphas at day $t$, which are aggregated into a composite alpha signal. This composite signal is then mapped to a target position $p_t \in [-1, 1]$, representing a normalized long or short exposure. Position sizing is further adjusted through a volatility-targeting mechanism to maintain a stable risk profile across different market conditions.

Trades are assumed to be executed immediately after the decision at day $t$, and the resulting position $p_t$ is held over the interval from day $t$ to day $t+1$. The realized strategy return is computed using the next-day return $r_{t+1}$, and the portfolio value is updated accordingly. Specifically, the strategy return at day $t+1$ is given by
\begin{equation}
R_{t+1} = p_t \cdot r_{t+1},
\end{equation}
where transaction costs and turnover effects are modeled separately during training.

This trading protocol ensures that signals are formed strictly prior to return realization, avoiding look-ahead bias and making the strategy fully implementable in practice.

\subsection{Evaluation}

\subsubsection{Baselines}

To evaluate the effectiveness of the proposed framework, we compare its trading performance with an equal-weighted strategy, buy-and-hold strategy, random entry/exit strategy, and momentum strategy. In the equal-weighted strategy, each LLM-generated alpha is assigned the same weight within the portfolio. The following provides a brief introduction to the remaining baseline strategies.

\paragraph{Buy and Hold (B\&H) Strategy} 
The Buy and Hold strategy serves as a passive benchmark representing the underlying asset's natural market growth. Following the logic in our implementation, the strategy enters a long position at the initial timestamp of the test period and maintains this exposure until the final step. This baseline provides a reference for the "Market Beta," allowing us to determine if the RL agent's active management adds value beyond simple market exposure.

\paragraph{Random Entry/Exit Strategy} 
To address the potential for stochastic success, we implement a Random Entry/Exit strategy that mimics the execution style of the PPO agent. This baseline is constrained by the empirical turnover and average position duration $D$ of the proposed PPO strategy. At each time step, the probability of a position switch is defined as $P(switch) = 1/D$. A state transition occurs if a randomly sampled value from a uniform distribution $u \sim U(0,1)$ is less than $P$. This ensures that the random baseline maintains a turnover rate and market exposure comparable to the PPO agent. To ensure statistical reliability, we conduct 1,000 Monte Carlo simulations and report the mean and standard deviation of the resulting metrics. This allows us to verify whether the agent's alpha is statistically distinct from a naïve, non-informative trading pattern.

\paragraph{Momentum (MOM) Strategy} 
The Momentum strategy serves as a technical baseline to evaluate whether the PPO agent can outperform a traditional trend-following heuristic. Based on the "momentum effect" observed in financial markets, this strategy calculates the return of each asset over a fixed look-back period of 10 days. The strategy enters a long position if the historical return is positive and a short position if it is negative. This baseline is essential for determining whether the RL agent is simply following prevailing price trends or if its integration of formulaic alphas enables it to identify more complex, non-linear market signals that a simple trend-following rule would miss.

\subsubsection{Evaluation Metrics}
To assess performance comprehensively, we use a range of metrics, including Information Coefficient (IC), Mutual Information, LightGBM Feature Importance, Cumulative Return, Maximum Drawdown, and Sharpe Ratio.

First, the \textbf{Information Coefficient (IC)} (Equation~\ref{eq:ic}) is calculated for each alpha signal across the ten selected companies. The IC measures the correlation between the predicted signals and the actual future returns. A positive IC value close to or above 0.05 indicates that the alpha has significant predictive power. IC values below 0.05 suggest weak predictive power, while negative IC values imply an inverse relationship, which may still be useful to construct contrarian strategies.

A contrarian strategy is an investment approach that opposes prevailing market trends. Contrarian investors buy when others are selling and sell when others are buying, anticipating price reversals~\cite{dreman1998contrarian}. In this context, a negative IC indicates that the signal moves in the opposite direction of return. Although the signal does not directly predict returns, it can still be valuable for contrarian strategies, where trading against the signal's direction is intentional.

\begin{equation}
IC = \frac{\text{Cov}(\text{rank}(\hat{r}), \text{rank}(r))}{\sigma_{\text{rank}(\hat{r})} \cdot \sigma_{\text{rank}(r)}}
\label{eq:ic}
\end{equation}

\noindent where:
\begin{itemize}
    \item $\hat{r}$ is the vector of predicted returns (alpha signals)
    \item $r$ is the vector of realized future returns
    \item $\text{rank}(\cdot)$ denotes the rank transformation (for Spearman correlation)
    \item $\text{Cov}(\cdot,\cdot)$ is the covariance between ranked variables
    \item $\sigma$ is the standard deviation of the ranked values
\end{itemize}

\textbf{Mutual Information (MI)} (Equation~\ref{eq: MI}) quantifies the amount of information shared between an input feature \( X \) and the target variable \( Y \). It measures how much knowing \( X \) reduces the uncertainty about \( Y \), capturing both linear and nonlinear relationships. Mathematically, the mutual information is defined as:

\begin{equation}
\label{eq: MI}
I(X; Y) = \sum_{x \in X} \sum_{y \in Y} p(x, y) \log \left( \frac{p(x, y)}{p(x)p(y)} \right)
\end{equation}

\noindent where \( p(x, y) \) is the joint probability distribution of \( X \) and \( Y \), and \( p(x) \), \( p(y) \) are their marginal distributions. A higher MI value indicates a stronger dependence between the feature and the target. Features with near-zero MI contribute little to predicting output and can often be discarded.

\textbf{LightGBM Feature Importance} (Equation~\ref{eq: FI}) is derived from the internal structure of the gradient boosting decision tree model. During training, LightGBM builds an ensemble of decision trees by sequentially minimizing a specified loss function. At each decision node, the model selects the feature that provides the greatest reduction in loss, commonly measured by information gain. The importance of a feature is then computed by summing its contribution to the loss reduction across all trees and relevant splits. This is formalized as follows.

\begin{equation}
\label{eq: FI}
\text{Importance}(f_i) = \sum_{t=1}^{T} \sum_{s \in \mathcal{S}_t(f_i)} \Delta \mathcal{L}_{t, s}
\end{equation}

\noindent where \( T \) is the total number of trees, \( \mathcal{S}_t(f_i) \) denotes the set of splits in tree \( t \) that involve feature \( f_i \), and \( \Delta \mathcal{L}_{t, s} \) represents the reduction in the loss at split \( s \).

LightGBM provides several types of feature importance metrics:

\begin{itemize}
    \item \textbf{Gain Importance:} Measures the total loss reduction brought by each feature across all splits. It reflects how much each feature contributes to the model’s performance and is considered the most informative metric.
    \item \textbf{Split Importance:} Counts the number of times a feature is used in splits, regardless of the magnitude of improvement. While it captures feature frequency, it may not accurately reflect predictive value.
    \item \textbf{Cover Importance:} Measures the number of samples affected by splits involving each feature. It helps assess how broadly a feature influences the dataset.
\end{itemize}

In this study, only the \textbf{Gain Importance} is reported and visualized for the LLM-generated alphas. This choice is motivated by the fact that gain importance directly quantifies a feature’s contribution to reducing prediction error. Since the primary objective is to identify which alphas most significantly improve model performance, gain importance offers a more relevant and interpretable measure. Other metrics, such as split and cover, might be influenced by frequent but weak splits or uneven sample distributions, making them less suitable for evaluating predictive strength.

The \textbf{Cumulative Return} measures the overall profitability of the trading strategy by aggregating returns over the entire evaluation period, reflecting the total gain or loss achieved through sequential trades.

\begin{equation}
R_{\text{cum}} = \prod_{t=1}^{T} (1 + r_t) - 1
\label{eq:cum_return}
\end{equation}

\noindent where:
\begin{itemize}
    \item $R_{\text{cum}}$ is the cumulative return over the time horizon $T$
    \item $r_t$ is the return at time $t$
    \item $\prod$ denotes the product over all time periods from $t = 1$ to $T$
\end{itemize}

In addition to cumulative return, we employ the \textbf{Sharpe Ratio} to evaluate the risk-adjusted performance of our PPO-adjusted trading strategy. The Sharpe Ratio measures the excess return per unit of risk and is defined as:
\begin{equation}
\text{Sharpe Ratio} = \frac{R_p - R_f}{\sigma_p}
\end{equation}
where \( R_p \) is the average return of the stock, \( R_f \) is the risk-free rate (typically the return of a short-term government bond), and \( \sigma_p \) is the standard deviation of stock returns. A higher Sharpe Ratio indicates that the strategy achieves better returns for each unit of risk taken. A Sharpe ratio higher than one is generally considered a sign of a good trading strategy, indicating that the strategy delivers returns significantly higher than the risk taken. This metric is particularly important when comparing strategies with different volatility profiles.

We also consider \textbf{Maximum Drawdown (MDD)} to evaluate downside risk. MDD represents the largest peak-to-trough decline in the value of a stock before a new peak is achieved. Formally, it is defined as:
\begin{equation}
\text{MDD} = \max_{t \in [0,T]} \left( \frac{\max_{\tau \in [0,t]} P(\tau) - P(t)}{\max_{\tau \in [0,t]} P(\tau)} \right)
\end{equation}
where \( P(t) \) denotes the stock value at time \( t \). Max Drawdown captures the worst historical loss from a stock high and is critical for understanding the potential downside exposure of a strategy. Lower values of MDD are preferable, indicating better capital preservation under adverse market conditions.

The combined use of these evaluation metrics provides a comprehensive view of both the predictive strength of the alphas and the practical profitability of the final strategy. Through evaluation, the effectiveness of the model in real-world trading scenarios can be more reliably assessed.

\subsubsection{Statistical Comparison via Diebold-Mariano Test}

To determine whether the performance gains of the PPO-based agent are statistically meaningful or merely a product of stochastic market noise, we utilize the Diebold-Mariano (DM) test. While traditional applications of the DM test focus on mean squared error in point forecasting, we adapt the framework to evaluate the relative profitability of trading trajectories.

We define the loss function $L(\cdot)$ as the negative daily net return of a strategy, such that $L(r_t) = -r_t$. The loss differential at time $t$ is then expressed as:
\begin{equation}
d_t = L(r_{t, \text{PPO}}) - L(r_{t, \text{Base}}) = r_{t, \text{Base}} - r_{t, \text{PPO}}
\end{equation}
where $r_{t, \text{PPO}}$ and $r_{t, \text{Base}}$ denote the net returns of the agent and the baseline, respectively. The null hypothesis ($H_0$) posits that there is no expected difference in the performance of the two strategies, effectively $\mathbb{E}[d_t] = 0$.

The DM test statistic is calculated as:
\begin{equation}
DM = \frac{\bar{d}}{\sqrt{\hat{\sigma}_d^2 / T}}
\end{equation}
where $\bar{d}$ represents the sample mean of the differential series and $\hat{\sigma}_d^2$ is a consistent estimate of the variance. Given that our trading decisions occur daily with a one-step horizon, we assume a forecast lag of $h=1$. A significantly negative $DM$ statistic accompanied by a $p$-value below the threshold of $\alpha = 0.05$ allows us to reject $H_0$, confirming that the agent's outperformance is statistically robust.

\subsubsection{Sharpe Ratio Bootstrap Test}

We evaluate the risk-adjusted performance difference between the PPO strategy and the Buy-and-Hold benchmark using a bootstrap-based Sharpe ratio test. Let $\{ r_t^{\text{PPO}} \}_{t=1}^{T}$ and $\{ r_t^{\text{B\&H}} \}_{t=1}^{T}$ denote the daily return series of the two strategies over the test period. For the PPO strategy, daily returns are obtained by averaging across multiple independent training runs with different random seeds.

The Sharpe ratio of strategy $s \in \{\text{PPO}, \text{B\&H}\}$ is defined as
\begin{equation}
SR_s = \frac{\mathbb{E}\!\left[r_t^{s}\right]}{\sqrt{\mathrm{Var}\!\left(r_t^{s}\right)}} .
\end{equation}
We focus on the relative Sharpe ratio difference,
\begin{equation}
\Delta SR = SR_{\text{PPO}} - SR_{\text{B\&H}} .
\end{equation}

To account for temporal dependence in financial returns, we adopt a block bootstrap procedure. The return series are resampled using contiguous blocks of fixed length, and for each bootstrap replication $b = 1, \dots, B$, a bootstrapped Sharpe difference $\Delta SR^{(b)}$ is computed. This yields an empirical distribution $\{ \Delta SR^{(b)} \}_{b=1}^{B}$.

A two-sided 95\% confidence interval is constructed from the empirical quantiles of the bootstrap distribution, and the corresponding $p$-value is obtained. Statistical significance is determined by whether the confidence interval excludes zero, indicating a meaningful difference in risk-adjusted performance between the PPO strategy and Buy-and-Hold.

\section{Results}\label{result}

\subsection{LLM-Generated Formulaic Alphas}

A sample of the LLM-Generated formulaic alphas is shown in Figure~\ref{tab:alpha_formulas}. These alphas are grouped by their main focus, including momentum, sentiment, volume, market indices, technical indicators, moving averages, volatility, and various combinations. These alphas incorporate price, volume, and sentiment features to capture diverse trading signals. 

Specifically, the momentum and moving average formulas (e.g., $\alpha_1, \alpha_{26}$) are based on the trend-following effect, where stock prices tend to keep moving in the same direction because investors often follow the crowd. Sentiment-based alphas (e.g., $\alpha_6, \alpha_{10}$) capture how investor emotions and news can push prices away from their true value in the short term. Volume and volatility formulas (e.g., $\alpha_{11}, \alpha_{36}$) use trading activity and price swings to measure market strength and risk. Additionally, the index-based and combination formulas show how local stocks are influenced by global markets, such as the S\&P 500 or Nikkei 225. By turning these ideas into simple formulas, the LLM allows the PPO agent to easily find and use the connections between price, sentiment, and global economic signals.

Table~\ref{tab:deepseek_thinking_process} shows a sample of the thinking process of the DeepSeek model. Unlike other LLMs, DeepSeek generates a thoughtful and detailed reasoning process before presenting results. This approach enhances the explainability of the outcomes, providing investors with clearer insights to inform their financial decisions. For example, when asked to create trading strategies, DeepSeek first analyzes the given features, such as stock prices, technical indicators, and sentiment scores, and identifies their roles in market behavior. It then thoughtfully designs each formula to capture different aspects like momentum, trend strength, volatility, and volume flow, making sure the strategies are diverse and aligned with typical trading logic. Only after carefully laying out this reasoning does DeepSeek present the final set of strategies, allowing users to clearly understand the rationale behind each output.

\begin{table}[h!]
\caption{LLM-generated Formulaic Alphas}
\label{tab:alpha_formulas}
\begin{tabular}{p{0.8\textwidth}}
\hline
\begin{tcolorbox}[colback=white, colframe=white, boxrule=0pt, sharp corners]
\scriptsize 
\# Momentum-based formulas \\
alpha1\_t = (C\_t - O\_t) / O\_t + 0.5 * Momentum\_3 \\
alpha2\_t = Momentum\_10 * (C\_t - SMA\_5) \\
alpha3\_t = (Momentum\_3 + Momentum\_10) / 2 \\
alpha4\_t = (C\_t - SMA\_20) * Momentum\_3 \\
alpha5\_t = Momentum\_3 * Momentum\_10 \\[1mm]

\# Sentiment-based formulas \\
alpha6\_t = S\_t * (C\_t - O\_t) / O\_t \\
alpha7\_t = Apple\_polarity * (C\_t - SMA\_5) \\
alpha8\_t = HSBC\_polarity * (C\_t - SMA\_20) \\
alpha9\_t = Pepsi\_polarity * Momentum\_3 \\
alpha10\_t = Tencent\_polarity * Momentum\_10 \\[1mm]

\# Volume-based formulas \\
alpha11\_t = V\_t / SMA\_20 \\
alpha12\_t = OBV * (C\_t - O\_t) / O\_t \\
alpha13\_t = V\_t * Momentum\_3 \\
alpha14\_t = V\_t * Momentum\_10 \\
alpha15\_t = OBV * S\_t \\[1mm]

\# Index-based formulas \\
alpha16\_t = (C\_t / Close\_Nikkei225) * S\_t \\
alpha17\_t = (C\_t / Close\_SP500) * Momentum\_3 \\
alpha18\_t = (C\_t / Close\_HSI) * Momentum\_10 \\
alpha19\_t = Close\_Nikkei225 * Close\_SP500 * Close\_HSI \\
alpha20\_t = (Close\_Nikkei225 + Close\_SP500 + Close\_HSI) / 3 \\[1mm]

\# Technical indicator-based formulas \\
alpha21\_t = MACD * MACD\_Signal \\
alpha22\_t = (MACD - MACD\_Signal) * S\_t \\
alpha23\_t = RSI\_14 * Momentum\_3 \\
alpha24\_t = RSI\_14 * Momentum\_10 \\
alpha25\_t = BB\_Upper - BB\_Lower \\[1mm]

\# Moving average-based formulas \\
alpha26\_t = (SMA\_5 - SMA\_20) * S\_t \\
alpha27\_t = (EMA\_10 - SMA\_20) * Momentum\_3 \\
alpha28\_t = SMA\_5 * SMA\_20 \\
alpha29\_t = EMA\_10 * SMA\_20 \\
alpha30\_t = (SMA\_5 + SMA\_20) / 2 \\[1mm]

\# Combination formulas \\
alpha31\_t = (C\_t - SMA\_5) / SMA\_5 + 0.5 * S\_t \\
alpha32\_t = (C\_t - SMA\_20) / SMA\_20 + 0.5 * Momentum\_3 \\
alpha33\_t = (C\_t - EMA\_10) / EMA\_10 + 0.5 * Momentum\_10 \\
alpha34\_t = (C\_t - SMA\_5) / SMA\_5 + (C\_t - SMA\_20) / SMA\_20 \\
alpha35\_t = (C\_t - SMA\_5) / SMA\_5 + (C\_t - EMA\_10) / EMA\_10 \\[1mm]

\# Volatility-based formulas \\
alpha36\_t = (High\_t - Low\_t) / Close\_t * S\_t \\
alpha37\_t = (High\_t - Low\_t) / Close\_t * Momentum\_3 \\
alpha38\_t = (High\_t - Low\_t) / Close\_t * Momentum\_10 \\
alpha39\_t = BB\_Upper - BB\_Lower \\
alpha40\_t = (BB\_Upper - C\_t) / (BB\_Upper - BB\_Lower) \\[1mm]

\# More combination formulas \\
alpha41\_t = (C\_t - SMA\_5) / SMA\_5 + (C\_t - SMA\_20) / SMA\_20 + S\_t \\
alpha42\_t = (C\_t - SMA\_5) / SMA\_5 + (C\_t - EMA\_10) / EMA\_10 + Momentum\_3 \\
alpha43\_t = (C\_t - SMA\_5) / SMA\_5 + (C\_t - EMA\_10) / EMA\_10 + Momentum\_10 \\
alpha44\_t = (C\_t - SMA\_5) / SMA\_5 + (C\_t - SMA\_20) / SMA\_20 + Momentum\_3 \\
alpha45\_t = (C\_t - SMA\_5) / SMA\_5 + (C\_t - SMA\_20) / SMA\_20 + Momentum\_10 \\[1mm]

\# Advanced combination formulas \\
alpha46\_t = (C\_t - SMA\_5) / SMA\_5 + (C\_t - SMA\_20) / SMA\_20 + S\_t + Momentum\_3 \\
alpha47\_t = (C\_t - SMA\_5) / SMA\_5 + (C\_t - EMA\_10) / EMA\_10 + S\_t + Momentum\_3 \\
alpha48\_t = (C\_t - SMA\_5) / SMA\_5 + (C\_t - EMA\_10) / EMA\_10 + S\_t + Momentum\_10 \\
alpha49\_t = (C\_t - SMA\_5) / SMA\_5 + (C\_t - SMA\_20) / SMA\_20 + S\_t + Momentum\_10 \\
alpha50\_t = (C\_t - SMA\_5) / SMA\_5 + (C\_t - SMA\_20) / SMA\_20 + S\_t + Momentum\_3 + Momentum\_10
\end{tcolorbox} \\
\hline
\end{tabular}
\end{table}

\begin{figure*}[!htbp]
  \centering
  \begin{subfigure}{\textwidth}
    \centering
    \includegraphics[width=\linewidth]{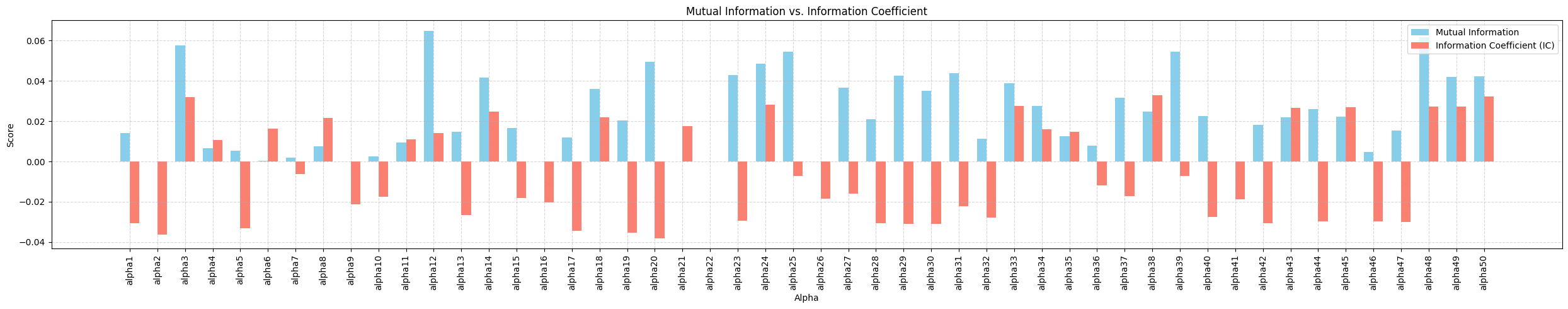}
    \caption{Apple}
  \end{subfigure}
  \hfill
  \begin{subfigure}{\textwidth}
    \centering
    \includegraphics[width=\linewidth]{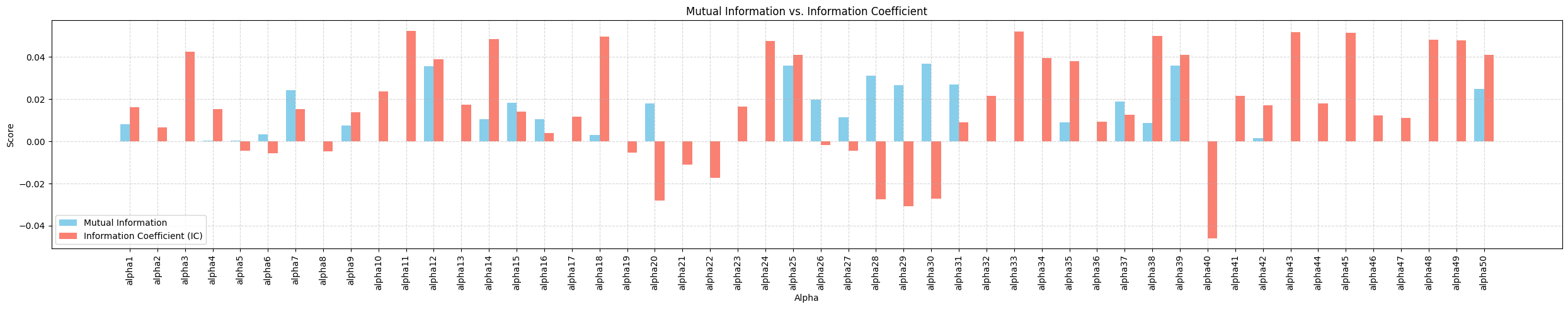}
    \caption{HSBC}
  \end{subfigure}

  \begin{subfigure}{\textwidth}
    \centering
    \includegraphics[width=\linewidth]{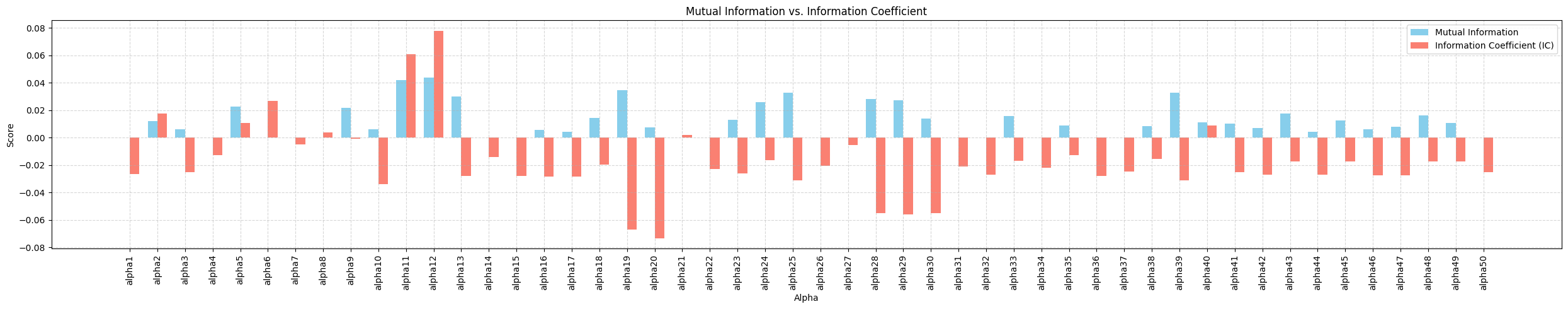}
    \caption{Tencent}
  \end{subfigure}
  \hfill
  \begin{subfigure}{\textwidth}
    \centering
    \includegraphics[width=\linewidth]{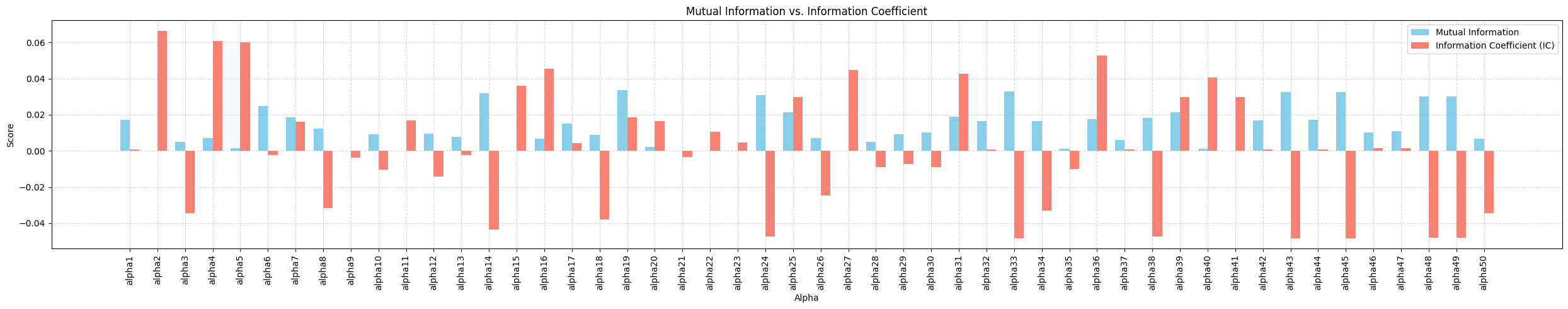}
    \caption{Toyota}
  \end{subfigure}

  \begin{subfigure}{\textwidth}
    \centering
    \includegraphics[width=\linewidth]{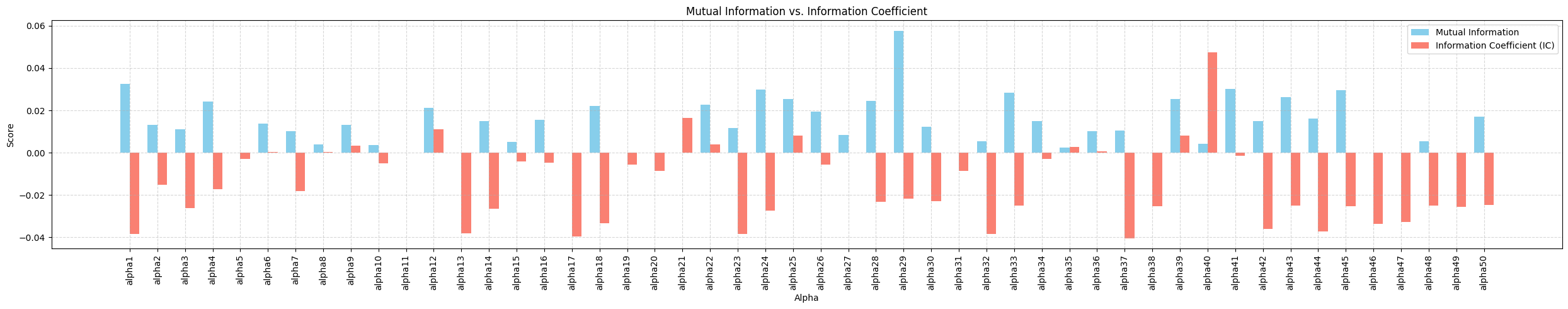}
    \caption{Pepsi}
  \end{subfigure}

  \caption{Mutual Information (in Blue) and Information Coefficient (in Red) of LLM-generated Alphas}
  \label{fig:MI_IC}
\end{figure*}

\begin{figure*}[!htbp]
  \centering
  \begin{subfigure}{0.3\textwidth}
    \centering
    \includegraphics[width=\linewidth]{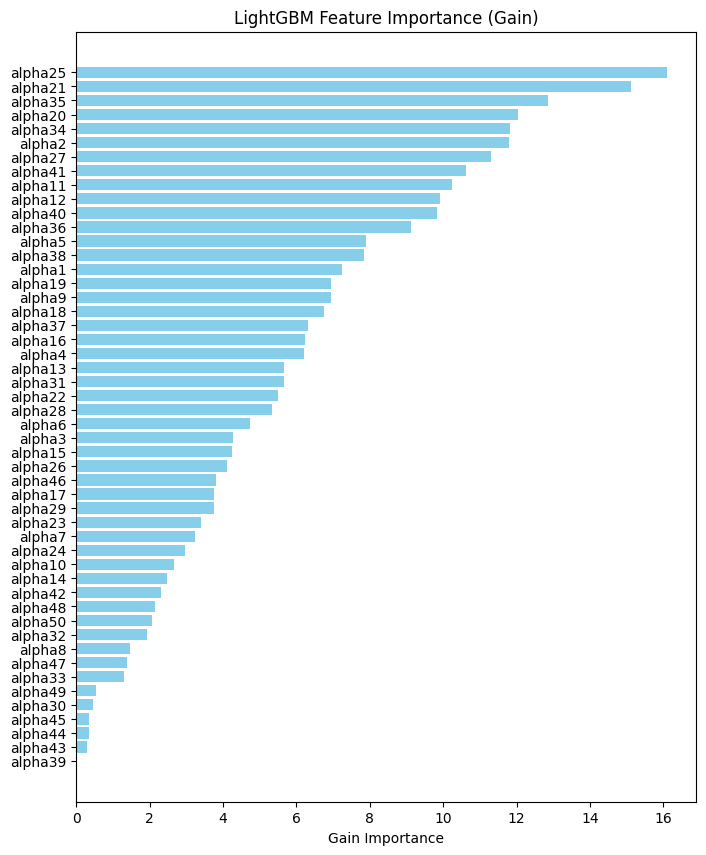}
    \caption{Apple}
  \end{subfigure}
  \hfill
  \begin{subfigure}{0.3\textwidth}
    \centering
    \includegraphics[width=\linewidth]{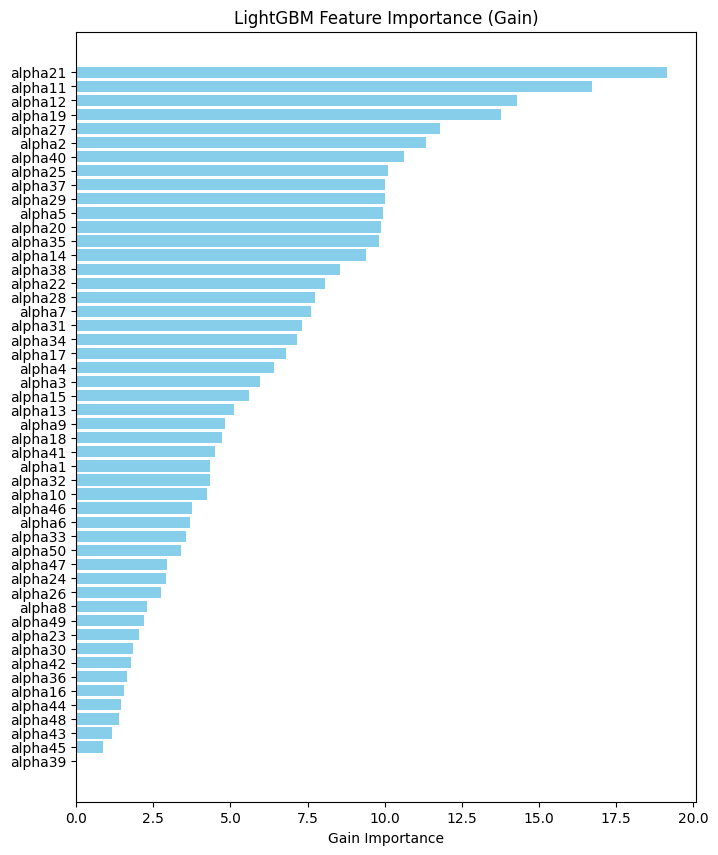}
    \caption{HSBC}
  \end{subfigure}

  \begin{subfigure}{0.3\textwidth}
    \centering
    \includegraphics[width=\linewidth]{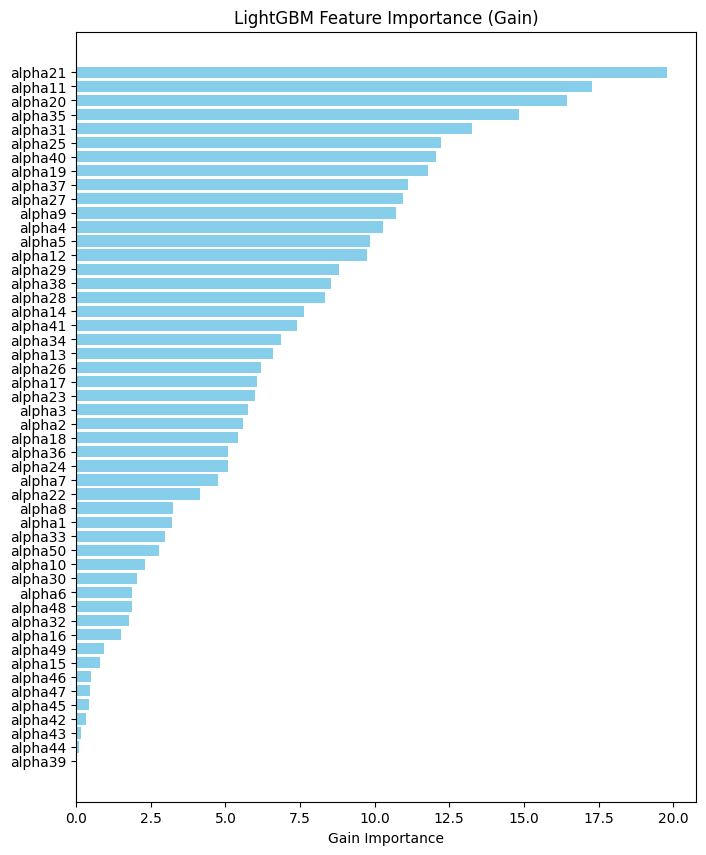}
    \caption{Tencent}
  \end{subfigure}
  \hfill
  \begin{subfigure}{0.3\textwidth}
    \centering
    \includegraphics[width=\linewidth]{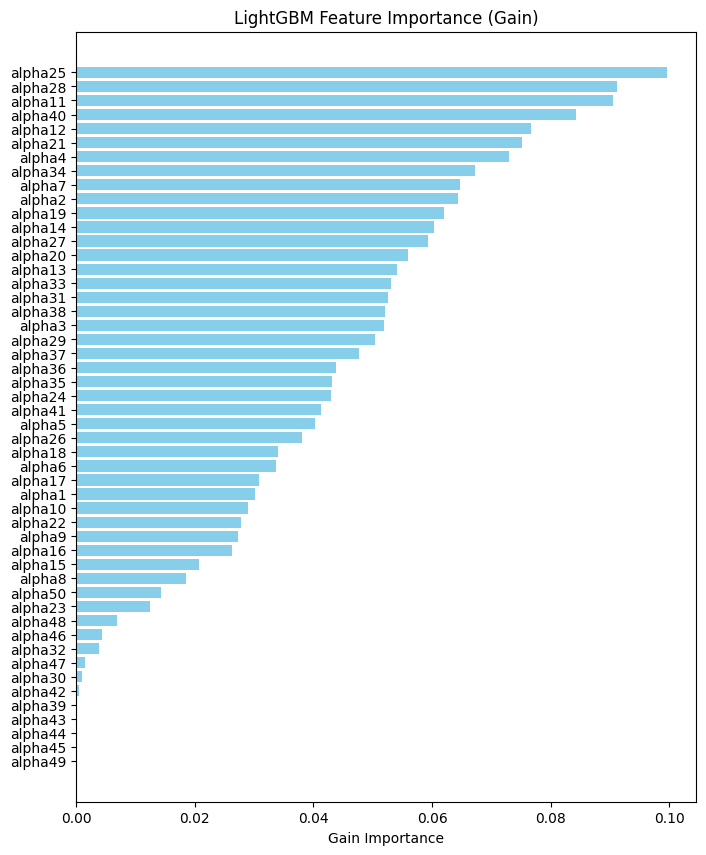}
    \caption{Toyota}
  \end{subfigure}

  \begin{subfigure}{0.3\textwidth}
    \centering
    \includegraphics[width=\linewidth]{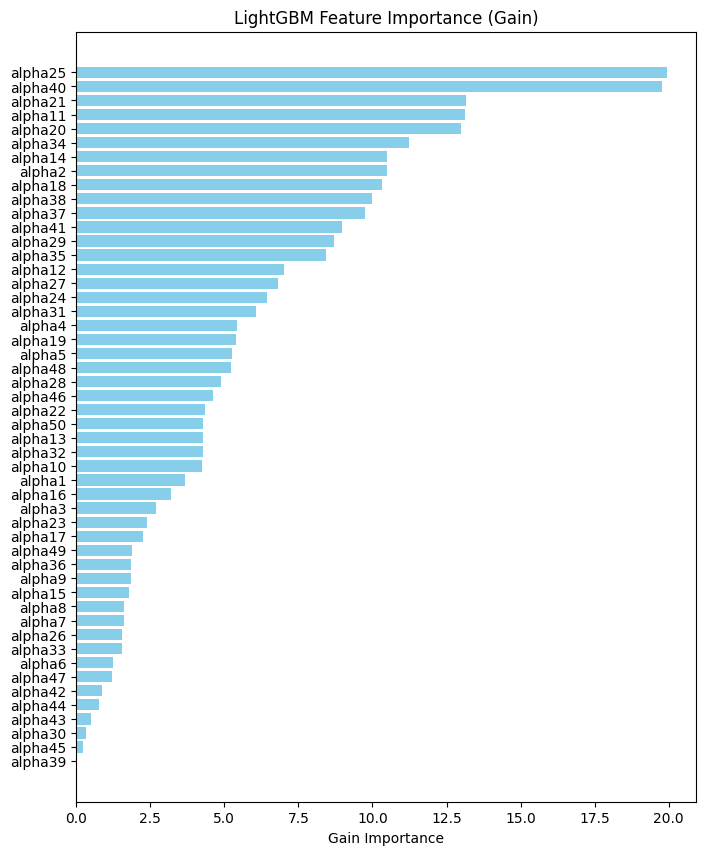}
    \caption{Pepsi}
  \end{subfigure}

  \caption{Feature Importance of LLM-generated Alphas}
  \label{fig:FI}
\end{figure*}

\begin{table*}[h!]
\centering
\caption{Deepseek Thinking Process}
\label{tab:deepseek_thinking_process}

\begin{tcolorbox}[
    colback=white,
    colframe=black,
    boxrule=0.5pt,
    sharp corners,
    width=\textwidth,
    enlarge left by=0mm,
    enlarge right by=0mm,
    halign=justify
]
\footnotesize
Alright, the user wants me to generate 50 alpha formulas using the given stock features. They specified that the formulas should be mathematical expressions combining these features and returned in Python syntax with specific variable names. 

First, I need to understand what each feature represents. Features like Open, High, Low, Close, Volume are standard OHLCV data. Then there are indices like Nikkei 225, S\&P 500, and HSI, which are market indicators. Sentiment and polarities for specific companies (Apple, HSBC, Pepsi, Tencent) are sentiment analysis metrics. The rest are technical indicators: SMA, EMA, Momentum, RSI, MACD, and OBV.

I should start by considering different categories of formulas. Maybe some based on momentum, others on volatility, sentiment, volume, etc. This way, the formulas will cover a broad range of strategies.

For momentum-based formulas, I can use the Momentum indicators, maybe combine them with price changes. For example, (C\_t - O\_t) / O\_t multiplied by Momentum\_3. That makes sense because it ties the day's price movement to recent momentum.

Next, I can think about volatility. Using High and Low prices, I can calculate daily ranges. Maybe (High\_t - Low\_t) / Close\_t to get a volatility measure. Then combine this with moving averages or sentiment.

Sentiment is another key area. The user provided overall sentiment and specific company polarities. I should incorporate these to capture market mood. For example, using S\_t (sentiment) multiplied by some function of price or volume.

Volume is important for confirmation. Formulas that include V\_t can indicate strength behind price movements. Maybe V\_t compared to a moving average of volume, or V\_t in relation to price changes.

I should also consider combinations of different indicators. For instance, using MACD and its signal line to generate a crossover signal, then scaling it by sentiment or volume.

Technical indicators like RSI can signal overbought or oversold conditions. I can use RSI levels to create mean-reversion strategies, perhaps combined with other indicators for confirmation.

I need to ensure each formula is unique and covers different aspects. Maybe some formulas focus on single indicators, while others combine multiple. For example, using BB\_Upper and BB\_Lower to create a volatility-based strategy, or combining SMA\_5 and SMA\_20 to look for crossovers.

I also need to think about the structure of each formula. They should be mathematical expressions, so I'll use operations like addition, subtraction, multiplication, division, and maybe exponents or logarithms where appropriate.

Let me start drafting some formulas. For the first one, a simple momentum strategy: (C\_t - O\_t) / O\_t. That's the daily return. Then, I can add a sentiment factor, maybe 0.5 * S\_t, to incorporate market mood.

Another idea: using the relative strength of the stock compared to an index. For example, (C\_t / Close\_Nikkei225) to see if the stock is outperforming the index.

I should also use technical indicators in creative ways. For example, MACD\_t - MACD\_Signal\_t gives the difference between the two lines, which can indicate a bullish or bearish signal.

Volume can be used to confirm the strength of a trend. So, V\_t / SMA\_Volume might show if current volume is higher than average, indicating strong movement.

I need to make sure I use all the provided features. That means including each feature in at least a few formulas. For example, using Apple\_polarity in some strategies, HSBC\_polarity in others, etc.

I can also create formulas that normalize the data, like (C\_t - SMA\_5) / SMA\_20, to see how the close price relates to different moving averages.

For each formula, I'll assign a variable name like alpha1\_t, alpha2\_t, etc., and express it in Python syntax using the specified variable abbreviations.

I should also vary the complexity. Some formulas can be simple two-term expressions, while others might have three or more terms, possibly with coefficients to weight different factors.

I need to ensure that each formula is plausible and could be used in a real trading strategy. They should be based on common trading logics, like trend-following, mean-reversion, momentum, etc.

Let me think of some more. Using RSI in a formula, perhaps as 1 / RSI\_t to indicate oversold conditions. Or combining RSI with momentum for a dual approach.

I should also consider combinations of different moving averages, like EMA\_10 and SMA\_20, to create crossovers or compare their values.

For the sentiment variables, I can use them as direct multipliers or add them after scaling. For example, S\_t * (C\_t - O\_t) / O\_t to tie sentiment to price movement.

I need to make sure that all 50 formulas are unique and don't repeat the same logic. Each should offer a different perspective or combination of features.

Finally, I'll review each formula to ensure they're mathematically sound and use the correct variable names as per the user's instructions.
\end{tcolorbox}

\end{table*}

\subsection{MI and IC Analysis}

Figure~\ref{fig:MI_IC} presents the mutual information (MI) and information coefficient (IC) of fifty LLM-generated alphas for five selected companies, chosen from the full set of ten companies for illustration. In the figure, the blue bars indicate the MI values, while the red bars represent the IC values. Together, they offer insight into both the statistical dependence between the alphas and returns (MI), and the directional predictive power (IC).

Apple shows the strongest overall signal among the five companies, with a larger number of alphas exhibiting positive MI values. For example, Alpha 3, 12, 14, 20, 23, 24, 25, 29, and 31 all have MI values exceeding 0.04, suggesting a moderate to strong dependency between alpha signals and Apple returns. However, their IC values vary and several of them are negative. This indicates that while the alphas are related to future returns, they may not correctly predict the direction. In particular, Alpha 3, 24, 33, 38, 48, 49, and 50 have higher IC values, highlighting a few strong directional signals.

Pepsi also shows relatively high MI values for many alphas, which implies that these features contain some information about future returns. However, most IC values for Pepsi are negative, suggesting that although the signals are related to the returns, their predictive direction is often incorrect or weak.

In contrast, HSBC stands out for its high IC values. Alphas such as Alpha 11, 14, 18, 24, 33, and 38 achieve IC values above 0.04, indicating a strong and consistent predictive direction. This suggests that LLM-generated alphas are more effective in forecasting HSBC return movements compared to other stocks.

On the other hand, Tencent exhibits generally low MI values and mostly negative IC values. This implies that LLM-generated alphas show a weak dependence on Tencent’s stock returns and also fail to capture directional accuracy. 

Toyota also shows a few alphas with notably strong predictive signals. In particular, Alpha 2, 4, and 5 achieve IC values exceeding 0.06, indicating a strong and reliable correlation with future stock returns. These high IC values suggest that these alphas consistently capture the correct direction of Toyota’s price movements. However, the majority of the alphas for Toyota fall within the IC range of 0.01 to 0.03, which reflects relatively weak but potentially usable signals. This distribution implies that while only a few alphas offer strong directional accuracy, many still exhibit a mild level of predictive power.

In general, these results suggest that the predictive quality of LLM-generated alphas can vary significantly between companies and that high mutual information does not always translate into high directional predictive power.

\subsection{Feature Importance}

Figure~\ref{fig:FI} presents the LightGBM feature importance scores of each LLM-generated alpha for five selected companies, chosen from the full set of ten for illustrative purposes. The results indicate that Alpha 25 stands out as the most influential feature for Apple, Toyota, and Pepsi, suggesting that this alpha consistently contributes valuable information to the prediction models for these companies. In contrast, Alpha 21 holds the highest importance for HSBC and Tencent, implying that different alphas capture distinct patterns depending on the company. On the other hand, Alpha 39 appears to have minimal importance across all five companies, indicating limited predictive relevance in most cases. While there are differences in which alphas are prioritized, the overall distribution of importance scores does not show extreme variation, suggesting a relatively balanced reliance on multiple alphas rather than dominance by a few. This diversity highlights that the predictive power of each alpha may depend heavily on the underlying characteristics of each stock.

\subsection{Comparison of PPO-Adjusted Strategy and Benchmarks}

\paragraph{Trading performance comparison}

Table~\ref{tab:ppo_vs_benchmarks} presents the performance metrics of both the PPO-adjusted strategy and different benchmarks. For each stock, the PPO framework is executed ten times using non-deterministic inference to allow stochastic behavior, which is an inherent feature of the PPO adjustment and important for capturing the variability in trading outcomes. The results are reported as the mean and standard deviation in the table. Reporting both statistics provides a measure of the average performance as well as the variability across runs, thus reflecting the stability and robustness of the framework.

Overall, the PPO strategy delivers more favorable risk-adjusted outcomes across most stocks in terms of Sharpe Ratio. Although the Buy-and-Hold (B\&H) strategy achieves higher cumulative returns in most cases, such as Toyota, Airbus, and Netflix, these gains are accompanied by higher risk exposure. In contrast, the PPO strategy records consistently higher Sharpe Ratios for nearly all stocks, indicating a more stable return profile.

A notable feature of the PPO strategy is its strong control of downside risk. For all ten stocks, the maximum drawdown remains very small, often below $1\%$. This behavior suggests that the agent has learned to reduce exposure during unfavorable market conditions, including periods of elevated volatility or persistent price declines. By comparison, the Equal-Weighted (EW) and Momentum (MOM) strategies experience substantial drawdowns, exceeding $50\%$ in several cases, such as Apple and Tencent. These results highlight the PPO agent’s emphasis on capital preservation rather than aggressive return seeking.

\paragraph{Results of significance tests}

To examine whether the observed performance differences are statistically meaningful, Table~\ref{tab:dm_results} presents the Diebold--Mariano (DM) test results based on daily returns. In most comparisons, the DM statistics are negative, indicating lower forecast errors for the PPO strategy relative to the benchmark methods.

When compared with the weaker baselines, namely the EW and MOM strategies, the PPO strategy shows statistically significant improvements at the $1\%$ or $5\%$ level for most of the stocks. This indicates that the learned policy captures market information that simple heuristic rules fail to reflect.

In contrast, the DM test results against the B\&H strategy are less frequently significant. For example, the $p$-values for Apple ($p = 0.071$) and HSBC ($p = 0.224$) do not reach conventional significance levels. This outcome suggests that during extended upward market phases, the daily return patterns of the PPO strategy and B\&H can be similar, even though their risk characteristics differ. However, according to the Sharpe ratio bootstrap test result in Table~\ref{tab:sharpe_ratio_test_results}, the PPO strategy yields higher Sharpe ratios than Buy-and-Hold for most stocks. Statistically significant improvements are observed in seven out of ten cases, while the remaining stocks show no significant differences. These findings indicate that PPO offers consistent risk-adjusted performance gains, although the magnitude and significance of the improvement depend on the underlying asset.

\begin{figure*}[!htbp]
    \centering
    \includegraphics[width=1\textwidth]{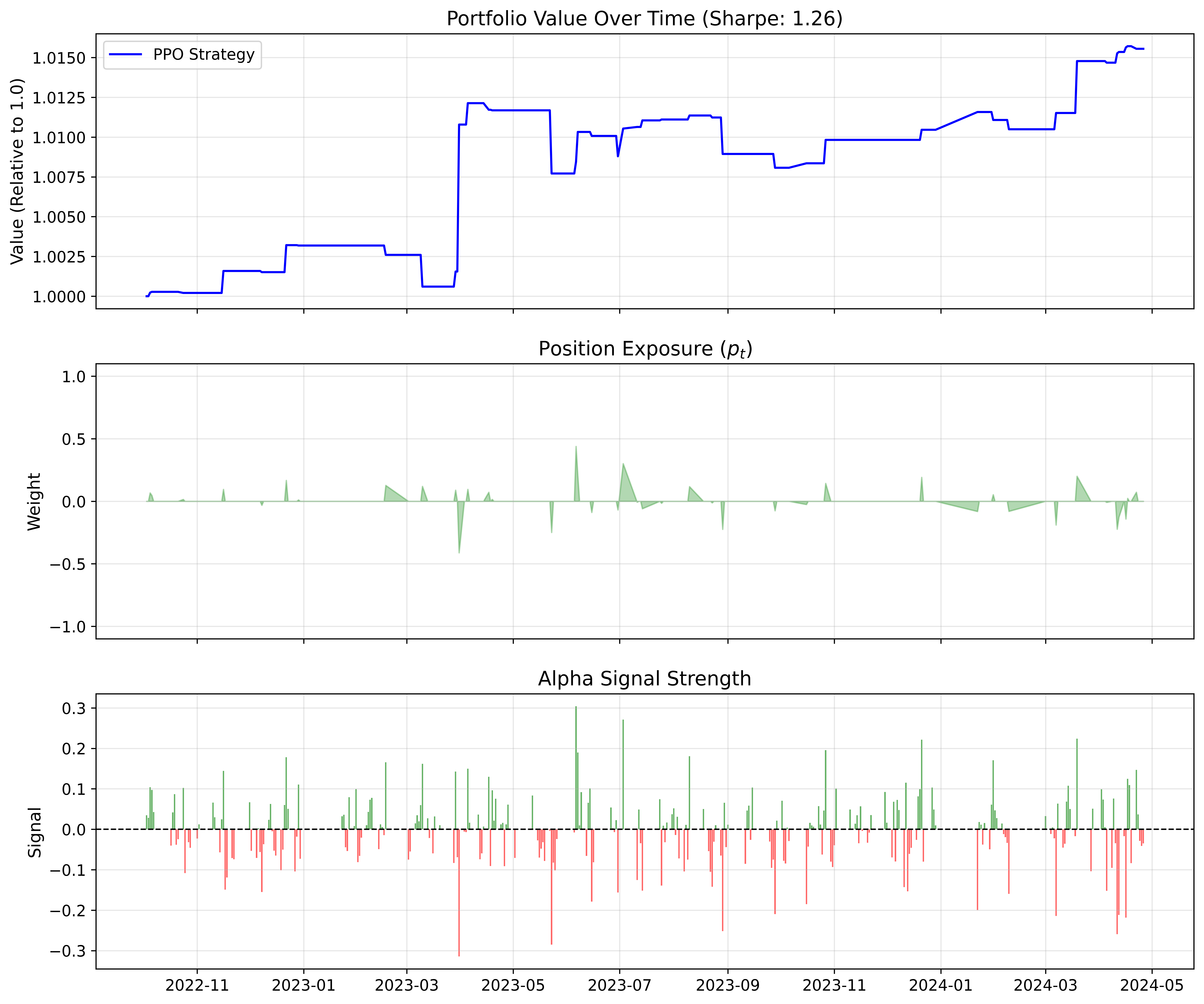}  % Adjust the width as needed
    \caption{Performance Analysis and Execution Dynamics of the PPO Strategy for Pepsi Stock. The panels illustrate the cumulative portfolio value (top), the volatility-adjusted position exposure $p_t$ (middle), and the composite alpha signal strength relative to decision thresholds (bottom).}
    \label{fig:PPO_Strategy}
\end{figure*}

A possible explanation for the lack of statistically significant DM test results when comparing the PPO strategy with the Buy-and-Hold (B\&H) benchmark relates to differences in trading activity. Figure~\ref{fig:PPO_Strategy} illustrates the portfolio value, position exposure, and alpha signal strength over the test period for Airbus. As is shown in Figure~\ref{fig:PPO_Strategy}, the PPO agent adopts a highly selective execution policy and frequently remains market-neutral for extended periods. As a result, a substantial portion of the return series consists of zero or near-zero values corresponding to no-trade days.

Since the Diebold--Mariano test assigns equal weight to all observations, including periods without active trading, strategies with sparse execution may experience a dilution effect in the test statistic. From a statistical perspective, this selective participation effectively reduces the informative sample size relative to continuously invested benchmarks such as B\&H. Consequently, the resulting $p$-values may be more conservative, potentially understating the statistical significance of the PPO agent’s predictive signal despite its superior risk-adjusted performance.

Finally, comparisons with the Random Entry/Exit Baseline (RB) show strong statistical evidence in favor of the PPO strategy for most stocks, including Toyota and Tencent (both with $p < 0.001$). This confirms that the observed performance is not driven by random trading behavior but instead reflects the effect of policy learning.

The consistency between the performance indicators in Table~\ref{tab:ppo_vs_benchmarks} and the DM test results in Table~\ref{tab:dm_results} points to a stable and reliable trading policy. Rather than maximizing raw returns, the PPO agent appears to focus on improving the Sharpe Ratio by limiting downside variability. Even in cases where cumulative returns fall below those of the B\&H strategy, such as Airbus, the substantially lower drawdowns lead to a more efficient risk--return trade-off. Overall, these findings suggest that reinforcement learning can be an effective tool for constructing trading strategies that balance profitability with robustness under volatile market conditions.

\begin{sidewaystable*}
\centering
\caption{Comprehensive Performance Analysis: PPO Strategy vs. EW, Buy \& Hold, Momentum, and Random Entry/Exit Baselines. The best result under each metric for each stock is highlighted in bold.}
\label{tab:ppo_vs_benchmarks}
\vspace{0.5cm}
\resizebox{\textheight}{!}{%
\begin{tabular}{|l|ccccc|ccccc|ccccc|}
\hline
\multirow{2}{*}{\textbf{Stock}} 
& \multicolumn{5}{c|}{\textbf{Cumulative Return}} 
& \multicolumn{5}{c|}{\textbf{Sharpe Ratio}} 
& \multicolumn{5}{c|}{\textbf{Max Drawdown}} \\ 
 & \textbf{PPO (Std)} & \textbf{EW} & \textbf{B\&H} & \textbf{MOM} & \textbf{RB (Std)} 
 & \textbf{PPO (Std)} & \textbf{EW} & \textbf{B\&H} & \textbf{MOM} & \textbf{RB (Std)} 
 & \textbf{PPO (Std)} & \textbf{EW} & \textbf{B\&H} & \textbf{MOM} & \textbf{RB (Std)} \\ \hline

\textbf{Apple} 
& \makecell{\textbf{1.6817} \\ (0.0619)} & $-0.3200$ & 0.3082 & $-0.4797$ & \makecell{0.1427 \\ (0.1640)}
& \makecell{\textbf{1.9998} \\ (0.0169)} & $-0.3601$ & 0.8194 & $-0.9119$ & \makecell{0.5283 \\ (0.5358)}
& \makecell{$\mathbf{-0.0101}$ \\ (0.0007)} & $-0.3663$ & $-0.1732$ & $-0.5214$ & \makecell{$-0.1726$ \\ (0.0484)} \\ \hline

\textbf{HSBC} 
& \makecell{0.5685 \\ (0.0259)} & $-0.9059$ & \textbf{0.6161} & $-0.5681$ & \makecell{0.2899 \\ (0.1671)}
& \makecell{\textbf{2.4113} \\ (0.0558)} & $-0.3580$ & 1.6087 & $-1.2032$ & \makecell{1.1838 \\ (0.5961)}
& \makecell{$\mathbf{-0.0220}$ \\ (0.0099)} & $-0.9489$ & $-0.1425$ & $-0.6479$ & \makecell{$-0.1179$ \\ (0.0333)} \\ \hline

\textbf{Pepsi} 
& \makecell{\textbf{0.6272} \\ (0.0331)} & $-0.1726$ & 0.1170 & 0.0231 & \makecell{0.0556 \\ (0.1012)}
& \makecell{\textbf{1.4319} \\ (0.0434)} & $-0.2879$ & 0.5220 & 0.2131 & \makecell{0.3292 \\ (0.5531)}
& \makecell{$\mathbf{-0.0067}$ \\ (0.0001)} & $-0.2922$ & $-0.1694$ & $-0.2698$ & \makecell{$-0.1305$ \\ (0.0423)} \\ \hline

\textbf{Tencent} 
& \makecell{\textbf{0.6245} \\ (0.0632)} & $-0.7074$ & 0.1752 & $-0.4757$ & \makecell{0.0999 \\ (0.2806)}
& \makecell{\textbf{1.1440} \\ (0.1052)} & $-0.2032$ & 0.4680 & $-0.5754$ & \makecell{0.2860 \\ (0.5614)}
& \makecell{$\mathbf{-0.0810}$ \\ (0.0208)} & $-0.8741$ & $-0.3362$ & $-0.6751$ & \makecell{$-0.2930$ \\ (0.0716)} \\ \hline

\textbf{Toyota} 
& \makecell{0.0638 \\ (0.0267)} & 0.1026 & \textbf{0.8742} & $-0.2464$ & \makecell{0.3956 \\ (0.2490)}
& \makecell{1.2786 \\ (0.3810)} & 0.1586 & \textbf{1.6646} & $-0.5608$ & \makecell{1.1780 \\ (0.6317)}
& \makecell{$\mathbf{-0.0223}$ \\ (0.0062)} & $-0.3571$ & $-0.1420$ & $-0.2820$ & \makecell{$-0.1422$ \\ (0.0365)} \\ \hline

\textbf{Airbus} 
& \makecell{ 0.0148 \\ (0.0079)} & $-0.3687$ & \textbf{0.6646} & $-0.1428$ & \makecell{ 0.2786 \\ (0.1269)}
& \makecell{1.3899 \\ (0.5449)} & $-0.1245$ & \textbf{2.4272} & $-0.6256$ & \makecell{ 1.6558 \\ (0.6503)}
& \makecell{ $\mathbf{-0.0061}$\\ (0.0042)} & $-0.5228$ & $-0.0757$ & $-0.2901$ & \makecell{ $-0.0830$ \\ (0.0241)} \\ \hline

\textbf{Exxon Mobil} 
& \makecell{ 0.0307 \\ (0.0175)} & $-0.1131$ & \textbf{0.4597} & 0.0684 & \makecell{ 0.1765 \\ (0.1374)}
& \makecell{ \textbf{1.5525} \\ (0.2590)} & $-0.0941$ & 1.5336 & 0.3741 & \makecell{0.9129 \\ (0.6201)}
& \makecell{ $\mathbf{-0.0030}$ \\ (0.0027)} & $-0.2994$ & $-0.1567$ & $-0.1358$ & \makecell{ $-0.1455$\\ (0.0388)} \\ \hline

\textbf{Petrobras} 
& \makecell{ 0.0945 \\ (0.0356)} & $-0.1403$ & \textbf{0.4659} & $-0.4124$ & \makecell{ 0.2553\\ (0.2505)}
& \makecell{ \textbf{1.5357}\\ (0.2272)} & 0.0736 & 1.1112 & $-1.1008$ & \makecell{0.8617 \\ (0.7045)}
& \makecell{ $\mathbf{-0.0102}$\\ (0.0042)} & $-0.5745$ & $-0.3154$ & $-0.5565$ & \makecell{ $-0.2313$ \\ (0.0783)} \\ \hline
\textbf{Netflix} 
& \makecell{ 0.3085 \\ (0.0009)} & $-0.3756$ & \textbf{0.6215} & $-0.1572$ & \makecell{ 0.2389 \\ (0.2541)}
& \makecell{\textbf{1.9010} \\ (0.0050)} & $-0.3469$ & 1.3044 & $-0.2247$ & \makecell{0.7507 \\ (0.6150)}
& \makecell{$\mathbf{-0.049}$ \\ (0.0004)} & $-0.4525$ & $-0.2010$ & $-0.3524$& \makecell{ $-0.1865$\\ (0.0484)} \\ \hline

\textbf{InfuSystem} 
& \makecell{ 0.0202 \\ (0.0182)} & $-0.6989$ & \textbf{0.0912} & $-0.0169$& \makecell{ 0.0666\\ (0.2297)}
& \makecell{ \textbf{0.8970} \\ (0.6531)} & $-0.5680$ & 0.4121 & 0.2005 & \makecell{ 0.2967 \\ (0.5985)}
& \makecell{ $\mathbf{-0.0101}$ \\ (0.0045)} & $-0.7443$ & $-0.3405$ & $-0.3225$ & \makecell{ $-0.2583$ \\ (0.0751)} \\ \hline

\end{tabular}}
\end{sidewaystable*}

\begin{table}[ht]
    \centering
    \small % Adjusts font size to fit all 8 metric columns
    \caption{Diebold-Mariano (DM) Test Results: PPO Strategy vs. Benchmarks. The PPO results are derived from the average daily returns across multiple independent training runs (seeds).}
    \label{tab:dm_results}
    \begin{tabular}{l cc cc cc cc}
        \toprule
        \multirow{2}{*}{\textbf{Stock}} & \multicolumn{2}{c}{\textbf{PPO vs. B\&H}} & \multicolumn{2}{c}{\textbf{PPO vs. Random}} & \multicolumn{2}{c}{\textbf{PPO vs. EW}} & \multicolumn{2}{c}{\textbf{PPO vs. MOM}} \\
        \cmidrule(lr){2-3} \cmidrule(lr){4-5} \cmidrule(lr){6-7} \cmidrule(lr){8-9}
        & DM Stat & $p$-value & DM Stat & $p$-value & DM Stat & $p$-value & DM Stat & $p$-value \\
        \midrule
        Apple   & $-1.81$ & 0.071 & $-3.26$** & 0.001 & $-4.96$** & 0.000 & $-2.49$* & 0.013 \\
        HSBC    & $-1.22$  & 0.224 & $-2.46$* & 0.014 & $-2.43$* & 0.015 & $-2.86$* & 0.004\\
        Pepsi   & $-1.34$  & 0.180 & $-1.79$  & 0.074 & $-2.59$** & 0.009 & $-1.51$ & 0.131\\
        Tencent & $-2.25$* & 0.024 & $-3.49$**  & 0.000 & $-0.37$ & 0.710 & $-3.07$** & 0.002 \\
        Toyota  & $-1.34$  & 0.180 & $-3.53$**  & 0.000 & $-1.38$ & 0.168 & $-3.76$** & 0.000 \\
        Airbus  &  $-0.54$ & 0.589 & $-0.20$* & 0.045 & $-3.94$** & $0.001$ & $-1.74$ & 0.084 \\
        Exxon Mobil  &  0.49 & 0.625 & 0.40  & 0.689 & $-1.96$* & 0.050 & $-0.27$ & 0.783 \\
        Petrobras  & 0.14  & 0.888 &  $-1.78$ & 0.075 & $-2.73$** & 0.006 & $-2.05$* & 0.040 \\
        Netflix  & $-0.61$ & 0.546 & $-1.23$  & 0.222 & $-3.27$** & 0.002  & $-2.16$* & 0.032 \\
        InfuSystem  &  $-1.18$ & 0.240 & $-2.06$*  & 0.041 & $-5.27$** & 0.000 & $-0.43$ & 0.665 \\
        \bottomrule
        \addlinespace[1ex]
        \multicolumn{9}{l}{\small \textit{Note: * and ** indicate significance at the 5\% and 1\% levels, respectively.}} \\
        \multicolumn{9}{l}{\small \textit{A negative DM stat indicates PPO outperforms the benchmark.}}
    \end{tabular}
\end{table}

\begin{table}[ht]
    \centering
    \small
    \caption{Relative Sharpe Ratio Bootstrap Test Results (PPO vs. Buy-and-Hold). The Sharpe ratio difference is defined as $\Delta SR = SR_{\text{PPO}} - SR_{\text{B\&H}}$. PPO results are computed using average daily returns across multiple independent training runs (seeds).}
    \label{tab:sharpe_ratio_test_results}
    \begin{tabular}{l c c c}
        \toprule
        \textbf{Stock} 
        & \textbf{Sharpe Difference} 
        & \textbf{95\% CI} 
        & \textbf{$p$-value} \\
        \midrule
        Apple        & 1.7695* & [0.1787, 3.4424] & 0.0296 \\
        HSBC         &  0.7928*  &  $[0.4321, 2.2641]$ &   0.0006    \\
        Pepsi        &  0.3365* &    [0.0010, 1.9168]  & 0.0494      \\
        Tencent      &  1.3282 &   $[-0.6347, 3.1084]$  & 0.1820      \\
        Toyota       & 0.4890* & [0.2533, 1.2456] &  0.0010  \\
        Airbus       &  $-0.1380$  &   $[-2.9630, 2.2057]$               &  0.7814   \\
        Exxon Mobil  &  0.1628* &     [0.0857, 1.1749]   &  0.0370      \\
        Petrobras    &  0.0903  &  $[-1.3394, 0.6880]$ &  0.8062    \\
        Netflix      &  0.6016* &    [0.1274, 1.3747]   &   0.0174    \\
        InfuSystem   &   1.0333*     &    [0.5141, 2.4829]  &   $<0.0001$    \\
        \bottomrule
        \addlinespace[1ex]
        \multicolumn{4}{l}{\small \textit{Note: $^{*}$ indicates statistical significance at the 5\% level.}} \\

        \multicolumn{4}{l}{\small \textit{Note: The 95\% confidence intervals are obtained via block bootstrap resampling.}} \\
        \multicolumn{4}{l}{\small \textit{Statistical significance is determined by whether the confidence interval excludes zero.}} \\
        \multicolumn{4}{l}{\small \textit{A positive Sharpe difference indicates that the PPO strategy outperforms Buy-and-Hold.}}
    \end{tabular}
\end{table}

\paragraph{PPO execution metrics}

Table~\ref{tab:ppo_practical_metrics} summarizes the execution-related properties of the PPO strategy. These indicators provide useful context for understanding how the agent translates its learned policy into actual trading behavior.

Across the ten stocks, the observed win rates range from $14.16\%$ to $44.19\%$. Although these values are relatively low in absolute terms, they should be viewed together with the strong risk-adjusted performance and limited drawdowns reported in Table~\ref{tab:ppo_vs_benchmarks}. This pattern suggests that the PPO agent does not depend on a high frequency of profitable trades. Instead, it appears to limit losses on unfavorable positions while allowing successful trades to run, thereby achieving stable performance despite a low hit ratio.

The average duration of open positions differs substantially across stocks, indicating flexible timing behavior. For more volatile or growth-oriented stocks such as Netflix, the agent tends to hold positions for longer periods, with an average duration of $14.05$ days. In contrast, for assets like Airbus and Toyota, the holding periods are much shorter, typically around $1.0$ to $1.3$ days. This difference suggests that the PPO policy adapts its trading horizon to asset-specific dynamics rather than applying a uniform temporal rule.

Furthermore, the strategy exhibits moderate trading activity. Average daily turnover remains low for several stocks, including Netflix ($0.0090$) and HSBC ($0.0348$), and reaches its highest level for Airbus ($0.2007$). Such turnover characteristics are important in practice, as they reduce the likelihood that transaction costs and market impact substantially diminish the strategy’s realized returns.

\begin{table}[h!]
\centering
\caption{Practical Trading Metrics Sample of the PPO-Adjusted Strategy across Ten Stocks}
\label{tab:ppo_practical_metrics}
\begin{tabular}{|l|ccc|}
\hline
\textbf{Stock} & \textbf{Avg. Win Rate} & \textbf{\makecell{Avg. Position \\ Duration (Days)}} & \textbf{Avg. Daily Turnover} \\ \hline
\textbf{Apple}   &   0.2582   &   3.8407   &   0.0616   \\ \hline
\textbf{HSBC}    &   0.3704   &   8.0426   &   0.0348   \\ \hline
\textbf{Pepsi}   &   0.1416   &   6.0783   &   0.0483   \\ \hline
\textbf{Tencent} &    0.1993   &   4.6610   &   0.0676      \\ \hline
\textbf{Toyota}  & 0.2911 & 1.3390 & 0.1015 \\ \hline
\textbf{Airbus}  & 0.4270 & 1.0001 & 0.2007 \\ \hline
\textbf{Exxon Mobil}  & 0.3783 & 1.0553 & 0.1795 \\ \hline
\textbf{Petrobras}  & 0.2360 & 1.6084 & 0.0989 \\ \hline
\textbf{Netflix}  & 0.4419 & 14.0526 & 0.0090 \\ \hline
\textbf{InfuSystem}  & 0.1729 & 1.8601 & 0.0613 \\ \hline
\end{tabular}
\end{table}

\section{Further Discussion and Analysis}

In this section, we select a subset of five stocks to conduct further analysis, such as the ablation analysis, wark-forward optimization analysis, impact of alpha selection settings, prompt information and sentiment on trading performance.

\subsection{Ablation Analysis of Alpha Sources}

To examine whether the use of a large language model contributes incremental value beyond conventional feature engineering, we conduct a controlled ablation analysis based on a subset of five stocks that contrasts LLM-generated alphas with a standardized set of human-crafted alphas. The latter consists of 101 manually designed factors~\cite{kakushadze2016101formulaicalphas} that are widely used in quantitative finance and serve as a representative benchmark for traditional alpha construction.

In order to ensure a fair comparison, the two alpha sources are evaluated under an identical experimental setup. Specifically, both sets are derived from the same underlying price and volume data, and share the same normalization procedures. Following this preprocessing stage, the retained alphas are fed into the same PPO-based optimization framework. All model architectures, hyperparameters, and reward definitions are kept fixed across experiments.

Since the number of LLM-generated alphas differs from the size of the human-crafted pool, we control for feature dimensionality by selecting a matched subset of 50 human-crafted alphas. This design choice prevents differences in performance from being driven by unequal feature counts and allows the comparison to focus on the alpha generation mechanism itself. Table~\ref{tab:human_alpha_examples} shows an example of five human-crafted alphas used for comparison.

By isolating the source of the alphas while holding all other components constant, this ablation study directly assesses whether the LLM provides additional predictive and economic value beyond established human-designed factors.

Table~\ref{tab:alpha_ablation_comparison} shows the performance comparison between human-crafted and LLM-generated alphas under the same PPO strategy. In general, the LLM-generated alphas have better performance for almost all stocks except Tencent, which suggests that LLM-generated alphas can serve as a reliable and effective alternative to manually designed signals within the PPO framework.

\begin{table}[htbp]
\centering
\small
\caption{Example of Human-Crafted Alphas Used for Comparison}
\label{tab:human_alpha_examples}
\begin{tabular}{c p{11cm}}
\toprule
\textbf{Alpha} & \textbf{Definition} \\
\midrule
Alpha 1 &
$\mathrm{Rank}\!\left(\mathrm{TsArgMax}\!\left(
\left(
\begin{cases}
\mathrm{Std}(\mathrm{Ret}, 20), & \text{if } \mathrm{Ret} < 0 \\
\mathrm{Close}, & \text{otherwise}
\end{cases}
\right)^2, 5
\right)\right)$ \\[6pt]

Alpha 2 &
$-\,\mathrm{Corr}\!\left(
\mathrm{Rank}\!\left(\Delta \log(\mathrm{Vol}), 2\right),
\mathrm{Rank}\!\left(\frac{\mathrm{Close} - \mathrm{Open}}{\mathrm{Open}}\right),
6
\right)$ \\[6pt]

Alpha 3 &
$-\,\mathrm{Corr}\!\left(
\mathrm{Rank}(\mathrm{Open}),
\mathrm{Rank}(\mathrm{Vol}),
10
\right)$ \\[6pt]

Alpha 4 &
$-\,\mathrm{TsRank}\!\left(\mathrm{Rank}(\mathrm{Low}), 9\right)$ \\[6pt]

Alpha 5 &
$\mathrm{Rank}\!\left(
\mathrm{Open} - \frac{1}{10}\sum_{t=1}^{10}\mathrm{VWAP}
\right)
\cdot
\left(
-\,\left|\mathrm{Rank}(\mathrm{Close} - \mathrm{VWAP})\right|
\right)$ \\
\bottomrule
\end{tabular}
\end{table}

\begin{table}[htbp]
\centering
\small
\caption{Comparison between Human-Crafted and LLM-Generated Alphas under the PPO Strategy for ten runs. Values are reported as Mean (Standard Deviation).}
\label{tab:alpha_ablation_comparison}

\begin{tabular}{l l cc}
\toprule
\textbf{Stock} & \textbf{Metric} & \textbf{Human-Crafted} & \textbf{LLM-Generated} \\
\midrule

\multirow{3}{*}{Apple}
& Cumulative Return & $-0.0085$ (0.0128) & 1.6817 (0.0619) \\
& Sharpe Ratio      & $-0.5914$ (0.9719) & 1.9998 (0.0169) \\
& Max Drawdown      & $-0.0125$ (0.0100) & $-0.0101$ (0.0007) \\
\midrule

\multirow{3}{*}{HSBC}
& Cumulative Return & 0.0040 (0.0074) & 0.5685 (0.0259) \\
& Sharpe Ratio      & 0.3402 (0.9468) & 2.4113 (0.0558) \\
& Max Drawdown      & $-0.0050$ (0.0041) & $-0.0220$ (0.0099) \\
\midrule

\multirow{3}{*}{Pepsi}
& Cumulative Return & $-0.0027$ (0.0313) & 0.6272 (0.0331) \\
& Sharpe Ratio      & $-0.0793$ (0.4538) & 1.4319 (0.0434) \\
& Max Drawdown      & $-0.0264$ (0.0095) & $-0.0067$ (0.0001) \\
\midrule

\multirow{3}{*}{Tencent}
& Cumulative Return & 0.0495 (0.0235) & 0.6245 (0.0632) \\
& Sharpe Ratio      & 2.0123 (0.5520) & 1.1440 (0.1052) \\
& Max Drawdown      & $-0.0051$ (0.0032) & $-0.0810$ (0.0208) \\
\midrule

\multirow{3}{*}{Toyota}
& Cumulative Return & $-0.0187$ (0.0285) & 0.0638 (0.0267) \\
& Sharpe Ratio      & $-0.5228$ (0.6423) & 1.2786 (0.3810) \\
& Max Drawdown      & $-0.0354$ (0.0249) & $-0.0223$ (0.0062) \\
\bottomrule
\end{tabular}

\end{table}

\subsection{Evaluation of different RL algorithms within the proposed framework}

In this subsection, we investigate whether PPO is the only model suitable for the proposed framework. We compare PPO with A2C, SAC, and TD3 under the same environmental setup. A brief overview of these algorithms is presented below.

\begin{itemize}
    \item \textbf{Advantage Actor--Critic (A2C)~\cite{mnih2016asynchronousmethodsdeepreinforcement}:} 
    A2C is a synchronous actor--critic method that updates the policy using on-policy samples. By estimating the advantage function, it reduces the variance of policy updates and provides a straightforward and stable baseline for policy-gradient-based trading strategies.

    \item \textbf{Soft Actor--Critic (SAC)~\cite{haarnoja2018softactorcriticoffpolicymaximum}:} 
    SAC is an off-policy algorithm that incorporates an entropy term into the objective function. This design encourages broader exploration and can lead to more stable learning, making SAC suitable for evaluating risk-aware and conservative trading behaviors under the same environment.

    \item \textbf{Twin Delayed Deep Deterministic Policy Gradient (TD3)~\cite{fujimoto2018addressingfunctionapproximationerror}:} 
    TD3 is an off-policy method that reduces value overestimation by using two critic networks and delaying policy updates. These mechanisms help stabilize learning in continuous action spaces, which is beneficial when controlling position sizes in trading tasks.
\end{itemize}

Table~\ref{tab:alpha_ablation_comparison} summarizes the performance of PPO and several baseline reinforcement learning algorithms across five selected stocks. Overall, the results indicate that model performance varies across both stocks and evaluation metrics, and no single method consistently achieves the best outcomes in all cases.

In terms of cumulative return, PPO tends to obtain higher values for Apple, Pepsi, and Tencent. However, SAC, TD3, and A2C often achieve higher Sharpe ratios for the same stocks. This pattern may suggest a trade-off between return and risk. A possible explanation is that PPO adopts relatively more active trading behaviors, which can increase return variability, while the other algorithms may favor more stable policies that lead to smoother returns. As a result, higher cumulative returns do not necessarily translate into higher risk-adjusted performance.

A notable difference is observed for Pepsi, where PPO produces a substantially larger cumulative return than the other methods. This gap may be related to the characteristics of the Pepsi stock during the evaluation period. It is possible that PPO is more sensitive to short-term signals or temporary trends, allowing it to capture certain profitable opportunities. In contrast, off-policy methods such as SAC and TD3 may converge to more conservative strategies, which could limit their exposure and reduce returns in this particular case.

In terms of downside risk, the maximum drawdown values are generally comparable across methods for most stocks. PPO exhibits slightly larger drawdowns in some cases, which is consistent with its higher return volatility. However, these differences remain relatively small and do not indicate excessive risk-taking.

Furthermore, the relatively small standard deviations observed across repeated experiments may be attributed to the risk-control mechanisms embedded in the trading environment. In particular, volatility targeting and bounded position sizing constrain the magnitude of portfolio exposure, which can limit performance variability across different runs. In addition, all models are evaluated on the same historical price trajectory under identical experimental settings, with differences arising mainly from policy initialization. Under such conditions, outcome variability is expected to remain limited.

In conclusion, this analysis suggests that PPO is not strictly required and may be substituted with other reinforcement learning models using the same environmental setup.

\begin{table}[!htbp]
    \centering
    \small
    \caption{Comparison between PPO and other RL algorithms. Values are reported as mean with standard deviation shown below.}
    \label{tab:alpha_ablation_comparison}
    \begin{tabularx}{\textwidth}{Xcccc}
        \toprule
        \textbf{Stock / Metric} 
        & \textbf{PPO} 
        & \textbf{SAC} 
        & \textbf{TD3} 
        & \textbf{A2C} \\
        \midrule
        
        \textbf{Apple} \\
        \quad Cumulative Return 
        & \makecell{1.6817 \\ (0.0619)} 
        & \makecell{0.5621 \\ (0.0075)} 
        & \makecell{0.5845 \\ (0.0085)} & \makecell{0.5619 \\ (0.0146)} \\
        \quad Sharpe Ratio      
        & \makecell{1.9998 \\ (0.0169)} 
        & \makecell{2.6061 \\ (0.0113)} 
        & \makecell{2.6236 \\ (0.0126)} & \makecell{2.5899 \\ (0.0266)} \\
        \quad Max Drawdown      
        & \makecell{$-0.0101$ \\ (0.0007)} 
        & \makecell{$-0.0099$ \\ (0.0003)} 
        & \makecell{$-0.0103$ \\ (0.0001)} & \makecell{$-0.0101$ \\ (0.0002)} \\
        \midrule
        
        \textbf{HSBC} \\
        \quad Cumulative Return 
        & \makecell{0.5685 \\ (0.0259)} 
        & \makecell{0.5563 \\ (0.0382)} 
        & \makecell{0.5605 \\ (0.0039)} & \makecell{0.5563 \\ (0.0145)} \\
        \quad Sharpe Ratio      
        & \makecell{2.4113 \\ (0.0558)} 
        & \makecell{2.3718 \\ (0.0958)} 
        & \makecell{2.3812 \\ (0.0113)} & \makecell{2.3746 \\ (0.0331)} \\
        \quad Max Drawdown      
        & \makecell{$-0.0220$ \\ (0.0007)} 
        & \makecell{$-0.0220$ \\ (0.0009)} 
        & \makecell{$-0.0220$ \\ (0.0001)} & \makecell{$-0.0220$ \\(0.0001)} \\
        \midrule
        
        \textbf{Pepsi} \\
        \quad Cumulative Return 
        & \makecell{0.6272 \\ (0.0331)} 
        & \makecell{0.0568 \\ (0.0003)} 
        & \makecell{0.0567 \\ (0.0002)} & \makecell{0.0569 \\ (0.0002)} \\
        \quad Sharpe Ratio      
        & \makecell{1.4319 \\ (0.0434)} 
        & \makecell{0.8597 \\ (0.0045)} 
        & \makecell{0.8587 \\ (0.0031)}& \makecell{0.8617 \\ (0.0030)} \\
        \quad Max Drawdown      
        & \makecell{$-0.0067$ \\ (0.0001)} 
        & \makecell{$-0.0010$ \\ (0.0001)} 
        & \makecell{$-0.0012$ \\ (0.0001)} & \makecell{$-0.0010$ \\ (0.0001)} \\
        \midrule
        
        \textbf{Tencent} \\
        \quad Cumulative Return 
        & \makecell{0.6245 \\ (0.0632)} 
        & \makecell{0.3270 \\ (0.0050)} 
        & \makecell{0.3262 \\ (0.0011)} & \makecell{0.3251 \\ (0.0021)} \\
        \quad Sharpe Ratio      
        & \makecell{1.1440 \\ (0.1052)} 
        & \makecell{1.8083 \\ (0.0216)} 
        & \makecell{1.8054 \\ (0.0051)} & \makecell{1.8011 \\ (0.0090)} \\
        \quad Max Drawdown      
        & \makecell{$-0.0810$ \\ (0.0208)} 
        & \makecell{$-0.0548$ \\ (0.0001)} 
        & \makecell{$-0.0541$ \\ (0.0003)} & \makecell{$-0.0541$ \\(0.0005)} \\
        \midrule
        
        \textbf{Toyota} \\
        \quad Cumulative Return 
        & \makecell{0.0638 \\ (0.0267)} 
        & \makecell{0.0722 \\ (0.0111)} 
        & \makecell{0.0579 \\ (0.0085)} & \makecell{0.0821 \\ (0.0265)} \\
        \quad Sharpe Ratio      
        & \makecell{1.2786 \\ (0.3810)} 
        & \makecell{1.2039 \\ (0.1694)} 
        & \makecell{1.2012 \\ (0.1665)} & \makecell{1.6646 \\ (0.3856)} \\
        \quad Max Drawdown      
        & \makecell{$-0.0223$ \\ (0.0062)} 
        & \makecell{$-0.0242$ \\ (0.0038)} 
        & \makecell{$-0.0211$ \\ (0.0028)} & \makecell{$-0.0158$ \\ (0.0032)} \\
        \bottomrule
    \end{tabularx}
\end{table}

\subsection{Walk-Forward Optimization Analysis} 
\label{sec: walk-forward analysis}

\begin{figure*}[!htbp]
    \centering
    \includegraphics[width=1\textwidth]{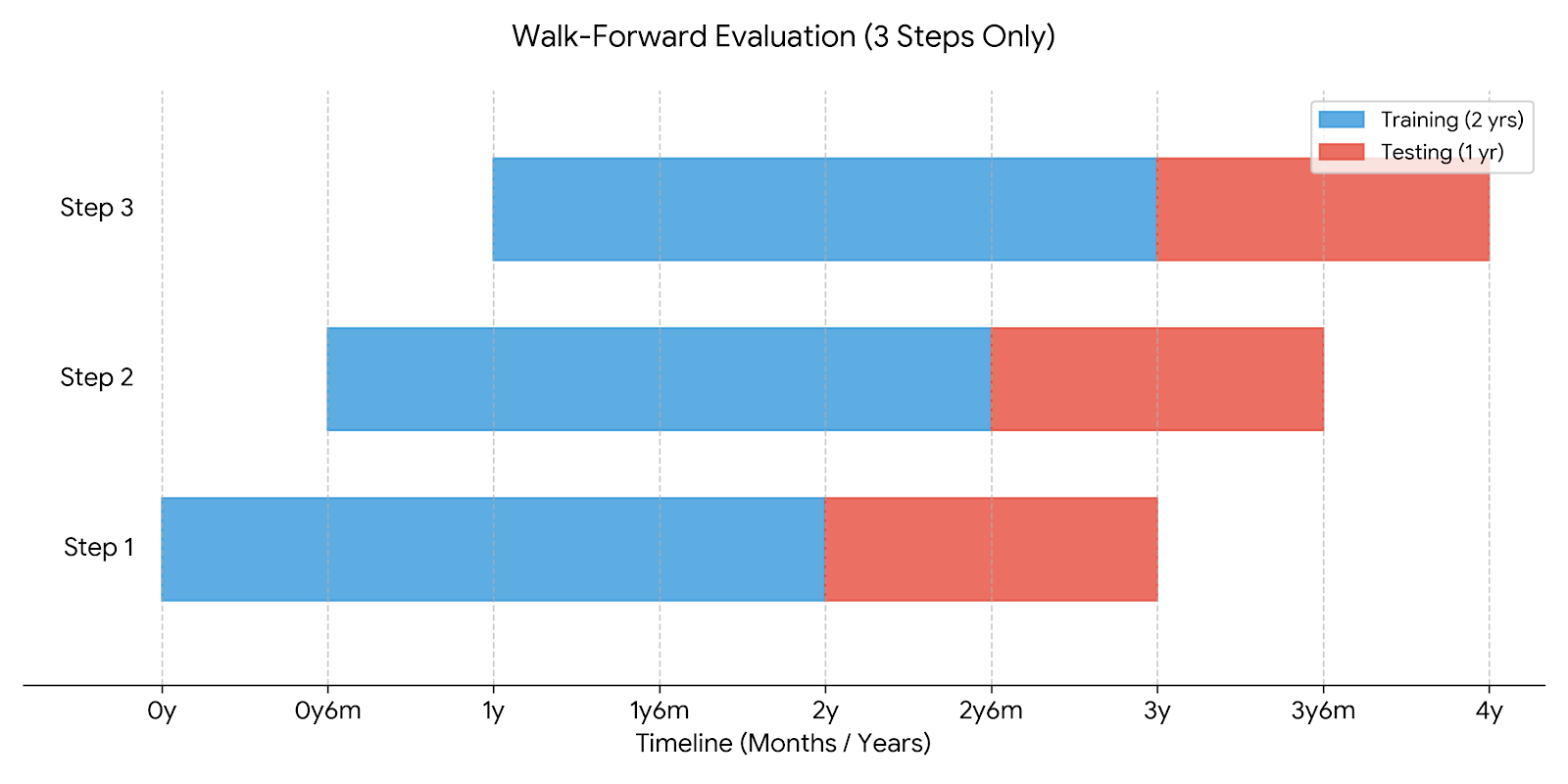}  % Adjust the width as needed
    \caption{Timeline illustration of walk-forward optimization. Each window consists of a rolling in-sample training period followed by an out-of-sample testing period. The window advances forward in time while maintaining a fixed training and testing length.}
    \label{fig:walk_forward}
\end{figure*}

In the original experiments, the dataset is split into a fixed training period and a subsequent testing period. Models are trained once on historical data and evaluated on unseen future data. To further evaluate the robustness of the proposed framework under different market conditions, we additionally conduct a walk-forward optimization analysis (Figure~\ref{fig:walk_forward}). Unlike the static train–test split, walk-forward optimization repeatedly retrains the model on a rolling training window and evaluates it on the immediately following out-of-sample period.

Financial markets exhibit strong non-stationarity and regime shifts. A single fixed split may overestimate performance if the training and testing periods share similar market conditions. Walk-forward optimization provides a more realistic evaluation by mimicking practical deployment, where models are periodically retrained using the most recent data. For walk-forward optimization, performance metrics are computed for each out-of-sample window and then averaged across all windows to obtain a stable estimate.

In this paper, we adopt a walk-forward evaluation scheme with a rolling training window of two years ($252\times2$ days), followed by a one-year (252 days) out-of-sample testing period. The window is advanced by six months (126 days) at each step, allowing partial overlap between consecutive test periods. This design aims to balance estimation stability and adaptability to evolving market conditions, while maintaining a strict separation between training and testing data. The PPO strategy is evaluated against a Buy-and-Hold strategy and a Momentum strategy within the same test window.

Table~\ref{tab:ppo_vs_bh_walk_forward} reports the walk-forward out-of-sample results for PPO, Buy-and-Hold, and a simple Momentum strategy across five stocks. Overall, the three methods show clearly different performance patterns under the same rolling evaluation setting.

Similar to the fixed train-test split scenario, across all stocks, Buy-and-Hold and Momentum strategies generally achieve higher total returns than PPO. This behavior is consistent with their continuous or near-continuous market exposure during the test windows. In contrast, PPO produces lower average returns, but it is associated with substantially lower annualized volatility and smaller maximum drawdowns. As a result, PPO often exhibits competitive, and in some cases higher, Sharpe ratios compared with the benchmark strategies.

The Momentum strategy behaves similarly to Buy-and-Hold for some stocks, such as Apple and Pepsi, where momentum signals remain positive for long periods. In these cases, the Momentum strategy maintains sustained exposure, leading to comparable returns, volatility, and drawdowns. For other stocks, such as HSBC and Toyota, Momentum shows weaker performance, which may indicate that trend persistence is less stable in these markets. However, this interpretation remains tentative, as the exact behavior depends on the underlying price dynamics within each test window.

The observed return–risk trade-off between PPO and the benchmark strategies appears reasonable in the walk-forward setting. Buy-and-Hold and Momentum strategies prioritize market exposure, which can lead to higher cumulative returns but also larger drawdowns and volatility. PPO, on the other hand, adopts a more cautious trading behavior, adjusting positions over time and limiting exposure during certain periods. This difference may help explain why PPO achieves lower drawdowns and volatility, even when raw returns are smaller.

The lower win rates observed for PPO do not necessarily imply inferior performance. A possible explanation is that PPO trades less frequently or remains inactive during many periods, which reduces the proportion of positive-return days but may help control risk. Similarly, the relatively small standard deviations reported for PPO across windows may be related to its more stable position adjustments.

Overall, the results suggest that PPO and traditional benchmark strategies exhibit distinct performance characteristics under walk-forward evaluation. PPO appears to emphasize stability and risk control, while Buy-and-Hold and Momentum focus more on sustained market exposure. These differences highlight the importance of evaluating trading strategies from both return and risk perspectives rather than relying on a single metric. Furthermore, the performance of PPO is relatively consistent with the fixed train-split result shown in Table~\ref{tab:ppo_vs_benchmarks}.

\begin{table}[htbp]
\centering
\small
\setlength{\tabcolsep}{4pt}
\caption{Walk-forward out-of-sample performance comparison between PPO and Buy-and-Hold.
Values are reported as mean (standard deviation) over all rolling test windows.}
\label{tab:ppo_vs_bh_walk_forward}
\begin{tabular}{@{}ll*{6}{c}@{}}
\toprule
\textbf{Stock} & \textbf{Method}
& \textbf{Total Return}
& \textbf{Ann. Vol.}
& \textbf{Sharpe}
& \textbf{Max DD}
& \textbf{Win Rate}
& \textbf{\makecell{Daily \\ Turnover}} \\
\midrule

\multirow{2}{*}{Apple}
& PPO
& \makecell{0.0479 \\ (0.0563)}
& \makecell{\textbf{0.0474} \\ (0.0520)}
& \makecell{\textbf{1.5530} \\ (0.8644)}
& \makecell{$\mathbf{-0.0040}$ \\ (0.0056)}
& \makecell{0.1392 \\ (0.1181)}
& \makecell{0.0427 \\ (0.0341)} \\
& Buy-and-Hold
& \makecell{0.3573 \\ (0.3912)}
& \makecell{0.3160 \\ (0.0837)}
& \makecell{0.9581 \\ (0.7357)}
& \makecell{$-0.2502$ \\ (0.0732)}
& \makecell{\textbf{0.5291} \\ (0.0293)}
& \makecell{0.0000 \\ (0.0000)} \\
& Momentum
& \makecell{\textbf{0.3665} \\ (0.4059)}
& \makecell{0.3156 \\ (0.0835)}
& \makecell{0.9684 \\ (0.7410)}
& \makecell{$-0.2530$ \\ (0.0695)}
& \makecell{0.5290 \\ (0.0296)}
& \makecell{1.000 \\ ($<0.0001$)} \\
\midrule

\multirow{2}{*}{HSBC}
& PPO
& \makecell{\textbf{0.0379} \\ (0.1219)}
& \makecell{\textbf{0.0913} \\ (0.2643)}
& \makecell{\textbf{1.1868} \\ (2.1042)}
& \makecell{$\mathbf{-0.0256}$ \\ (0.0510)}
& \makecell{0.1609 \\ (0.1398)}
& \makecell{0.0566 \\ (0.0559)} \\
& Buy-and-Hold
& \makecell{$-0.1855$ \\ (0.1814)}
& \makecell{0.2359 \\ (0.1070)}
& \makecell{$-0.7394$ \\ (0.8155)}
& \makecell{$-0.7082$ \\ (0.3647)}
& \makecell{\textbf{0.4843} \\ (0.0419)}
& \makecell{0.0000 \\ (0.0000)} \\
& Momentum
& \makecell{$-0.0600$ \\ (0.1437)}
& \makecell{0.0639 \\ (0.1030)}
& \makecell{$-0.8284$ \\ (0.9728)}
& \makecell{$-0.0941$ \\ (0.1660)}
& \makecell{0.1543 \\ (0.2322)}
& \makecell{0.3333 \\ (0.5000)} \\
\midrule

\multirow{2}{*}{Pepsi}
& PPO
& \makecell{0.0220 \\ (0.0657)}
& \makecell{\textbf{0.0389} \\ (0.0463)}
& \makecell{\textbf{0.8481} \\ (1.5084)}
& \makecell{$\mathbf{-0.0109}$ \\ (0.0291)}
& \makecell{0.1352 \\ (0.1435)}
& \makecell{0.0403 \\ (0.0394)} \\
& Buy-and-Hold
& \makecell{\textbf{0.1402} \\ (0.1187)}
& \makecell{0.2055 \\ (0.0849)}
& \makecell{0.8471 \\ (0.6614)}
& \makecell{$-0.1506$ \\ (0.0726)}
& \makecell{0.5430 \\ (0.0225)}
& \makecell{0.0000 \\ (0.0000)} \\
& Momentum
& \makecell{0.1398 \\ (0.1173)}
& \makecell{0.2054 \\ (0.0845)}
& \makecell{0.8409 \\ (0.6428)}
& \makecell{$-0.1506$ \\ (0.0726)}
& \makecell{\textbf{0.5433} \\ (0.0215)}
& \makecell{1.0000 \\ ($<0.0001$)} \\
\midrule

\multirow{2}{*}{Tencent}
& PPO
& \makecell{0.0097 \\ (0.0856)}
& \makecell{\textbf{0.0438} \\ (0.0514)}
& \makecell{0.6244 \\ (1.9088)}
& \makecell{$\mathbf{-0.0248}$ \\ (0.0497)}
& \makecell{0.1324 \\ (0.1117)}
& \makecell{0.0578 \\ (0.0579)} \\
& Buy-and-Hold
& \makecell{\textbf{0.2798} \\ (0.3753)}
& \makecell{0.3506 \\ (0.0719)}
& \makecell{0.8139 \\ (0.8226)}
& \makecell{$-0.6062$ \\ (0.3844)}
& \makecell{\textbf{0.4923} \\ (0.0337)}
& \makecell{0.0000 \\ (0.0000)} \\
& Momentum
& \makecell{0.1999 \\ (0.3224)}
& \makecell{0.2502 \\ (0.1807)}
& \makecell{\textbf{0.8200} \\ (0.7761)}
& \makecell{$-0.3840$ \\ (0.3812)}
& \makecell{0.3823 \\ (0.2193)}
& \makecell{0.7778 \\ (0.4410)} \\
\midrule

\multirow{2}{*}{Toyota}
& PPO
& \makecell{0.0098 \\ (0.0130)}
& \makecell{\textbf{0.0201} \\ (0.0133)}
& \makecell{\textbf{0.8924} \\ (1.0367)}
& \makecell{$\mathbf{-0.0070}$ \\ (0.0046)}
& \makecell{0.1024 \\ (0.0616)}
& \makecell{0.0644 \\ (0.0371)} \\
& Buy-and-Hold
& \makecell{\textbf{0.1704} \\ (0.2404)}
& \makecell{0.2472 \\ (0.0358)}
& \makecell{0.7302 \\ (0.8416)}
& \makecell{$-0.1911$ \\ (0.0565)}
& \makecell{\textbf{0.4984} \\ (0.0237)}
& \makecell{0.0000 \\ (0.0000)} \\
& Momentum
& \makecell{0.1203 \\ (0.2153)}
& \makecell{0.2232 \\ (0.0861)}
& \makecell{0.6022 \\ (0.7830)}
& \makecell{$-0.1779$ \\ (0.0817)}
& \makecell{0.4480 \\ (0.1591)}
& \makecell{0.9000 \\ (0.3162)} \\
\bottomrule
\end{tabular}
\end{table}

\subsection{Impact of Different Alpha Selection Settings on Trading Performance}

\begin{table*}[h]
\centering
\caption{Number of Alphas Selected by Different Reduction Methods for Each Company}
\label{tab:alpha_reduction}
\begin{tabular}{lccccc}
\toprule
\textbf{Method} & \textbf{Apple} & \textbf{HSBC} & \textbf{Pepsi} & \textbf{Tencent} & \textbf{Toyota} \\
\midrule
Low-Correlation Selection & 13 & 14 & 14 & 18 & 10 \\
High-Contribution Selection & 10 & 10 & 10 & 10 & 10 \\
Random Selection & 30 & 30 & 30 & 30 & 30 \\
\bottomrule
\end{tabular}
\end{table*}

\begin{sidewaystable}[ph!]
    \centering
    \renewcommand{\arraystretch}{1.2} % Adds a bit of vertical breathing room
    \caption{Comparison of PPO Strategies across different Alpha Selection Methods. The best value for each stock under each metric is bolded.}
    \label{tab:different_alpha_selection_method}
    
    \begin{tabularx}{\linewidth}{l XXXXX}
        \toprule
        \textbf{Metric / Alpha Strategy} & \textbf{Apple} & \textbf{HSBC} & \textbf{Pepsi} & \textbf{Tencent} & \textbf{Toyota} \\
        \midrule
        
        \textbf{Cumulative Return} \\
        \quad Low-Correlation   & 1.6641 (0.1755) & 0.4983 (0.0660) & \textbf{0.6319} (0.0775) & 0.5956 (0.0722) & \textbf{0.0842} (0.0579) \\
        \quad High-Contribution & 1.6072 (0.1093) & \textbf{0.5199} (0.0991) & 0.6023 (0.0441) & 0.5634 (0.1200) & 0.0741 (0.0395) \\
        \quad 30 Random         & \textbf{1.7276} (0.0647) & 0.4853 (0.0451) & 0.6176 (0.0225) & \textbf{0.6041} (0.0755) & 0.0410 (0.0291) \\
        \addlinespace
        
        \textbf{Sharpe Ratio} \\
        \quad Low-Correlation   & 2.0002 (0.0692) & 0.8388 (0.0776) & 1.4259 (0.0622) & \textbf{1.1191} (0.1018) & \textbf{0.2072} (0.1083) \\
        \quad High-Contribution & 1.9890 (0.0626) & \textbf{0.8791} (0.1215) & 1.3996 (0.0296) & 1.0620 (0.1735) & 0.1894 (0.0765) \\
        \quad 30 Random         & \textbf{2.0133} (0.0255) & 0.8399 (0.0534) & \textbf{1.4271} (0.0367) & 1.1129 (0.1132) & 0.1238 (0.0565) \\
        \addlinespace
        
        \textbf{Max Drawdown} \\
        \quad Low-Correlation   & $-0.0113$ (0.0016) & $\mathbf{-0.2210}$ (0.0121) & \textbf{$\mathbf{-0.0066}$} (0.0002) & $\mathbf{-0.0810}$ (0.0183) & $\mathbf{-0.2344}$ (0.0288) \\
        \quad High-Contribution & $-0.0113$ (0.0016) & $-0.2286$ (0.0253) & $-0.0076$ (0.0022) & $-0.0919$ (0.0263) & $-0.2625$ (0.0217) \\
        \quad 30 Random         & $\mathbf{-0.0109}$ (0.0009) & $-0.2248$ (0.0061) & $-0.0068$ (0.0006) & $-0.0937$ (0.0307) & $-0.2664$ (0.0137) \\
        \bottomrule
        \addlinespace[1ex]
        \multicolumn{6}{l}{\small \textit{Note: Values represent the mean across test runs with standard deviation in parentheses.}}
    \end{tabularx}
\end{sidewaystable}

\begin{figure*}[!htbp]
  \centering
  % First row
  \begin{subfigure}{0.45\textwidth}
    \centering
    \includegraphics[width=\linewidth]{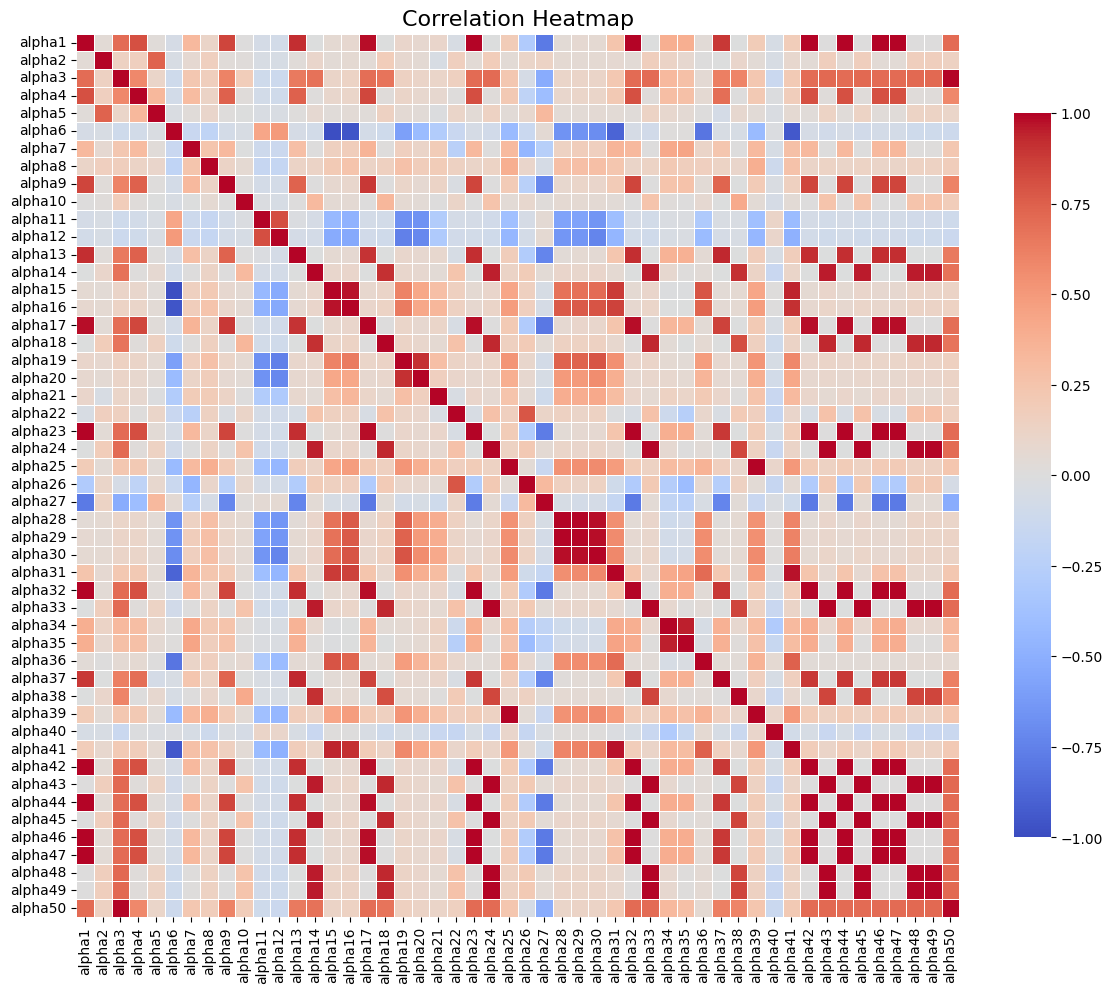}
    \caption{Apple}
  \end{subfigure}
  \hfill
  \begin{subfigure}{0.45\textwidth}
    \centering
    \includegraphics[width=\linewidth]{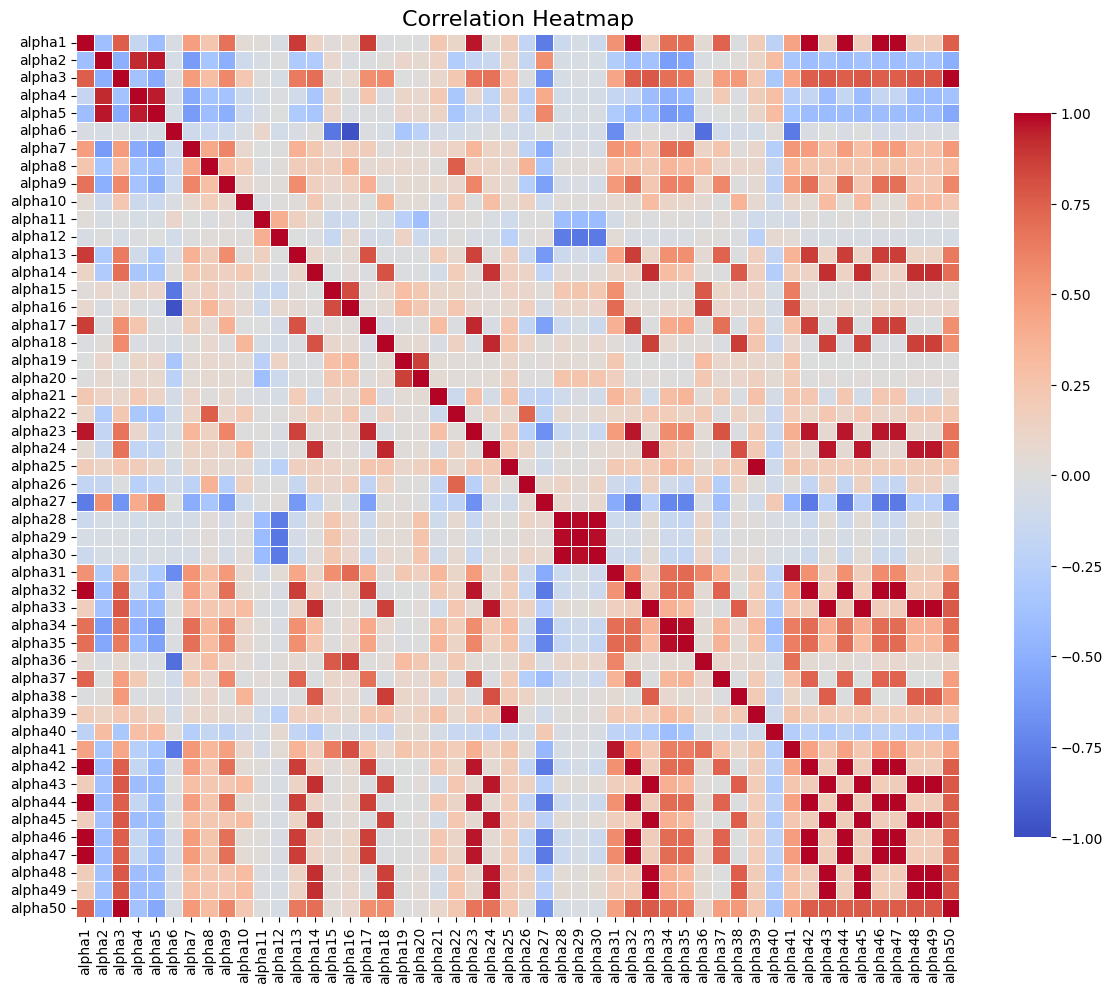}
    \caption{HSBC}
  \end{subfigure}
  \hfill
  \begin{subfigure}{0.45\textwidth}
    \centering
    \includegraphics[width=\linewidth]{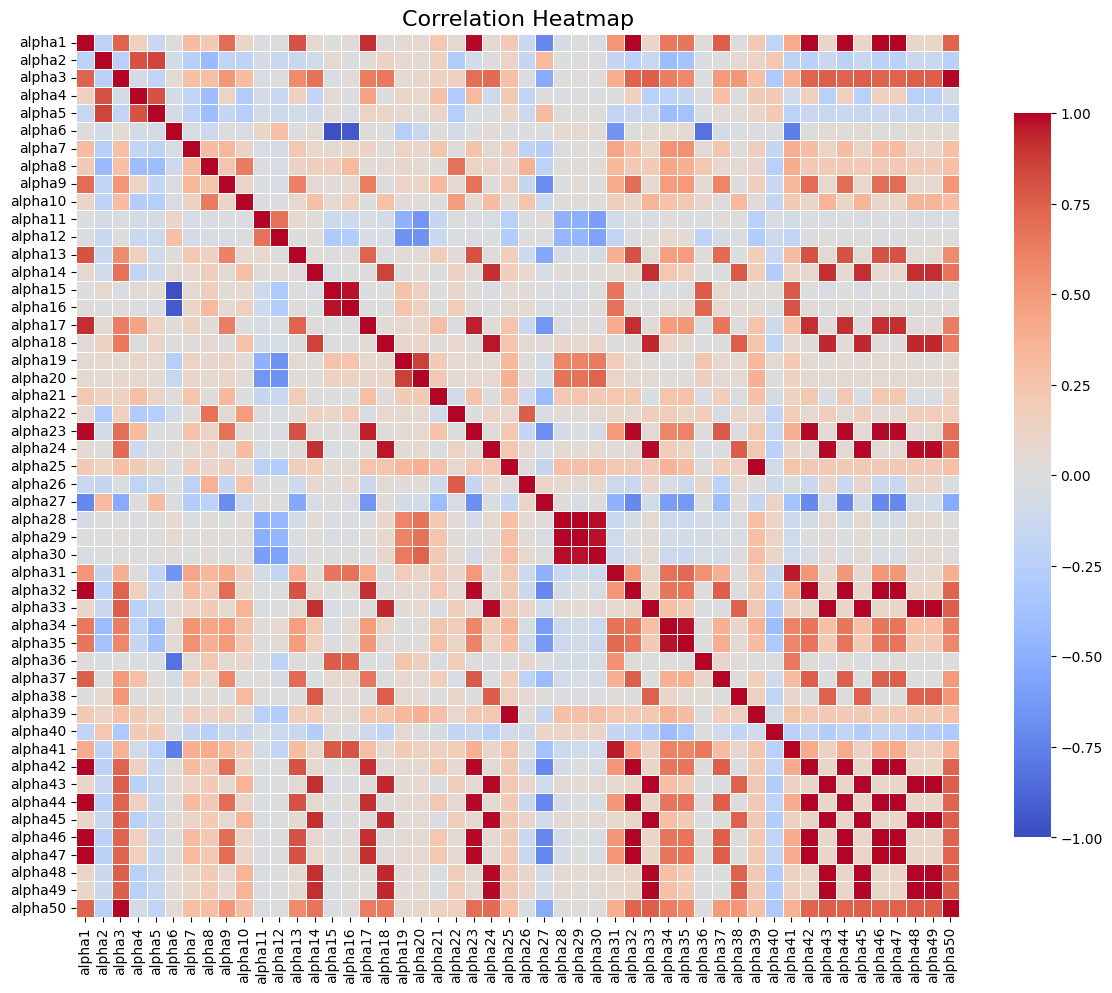}
    \caption{Tencent}
  \end{subfigure}
  \hfill
  \begin{subfigure}{0.45\textwidth}
    \centering
    \includegraphics[width=\linewidth]{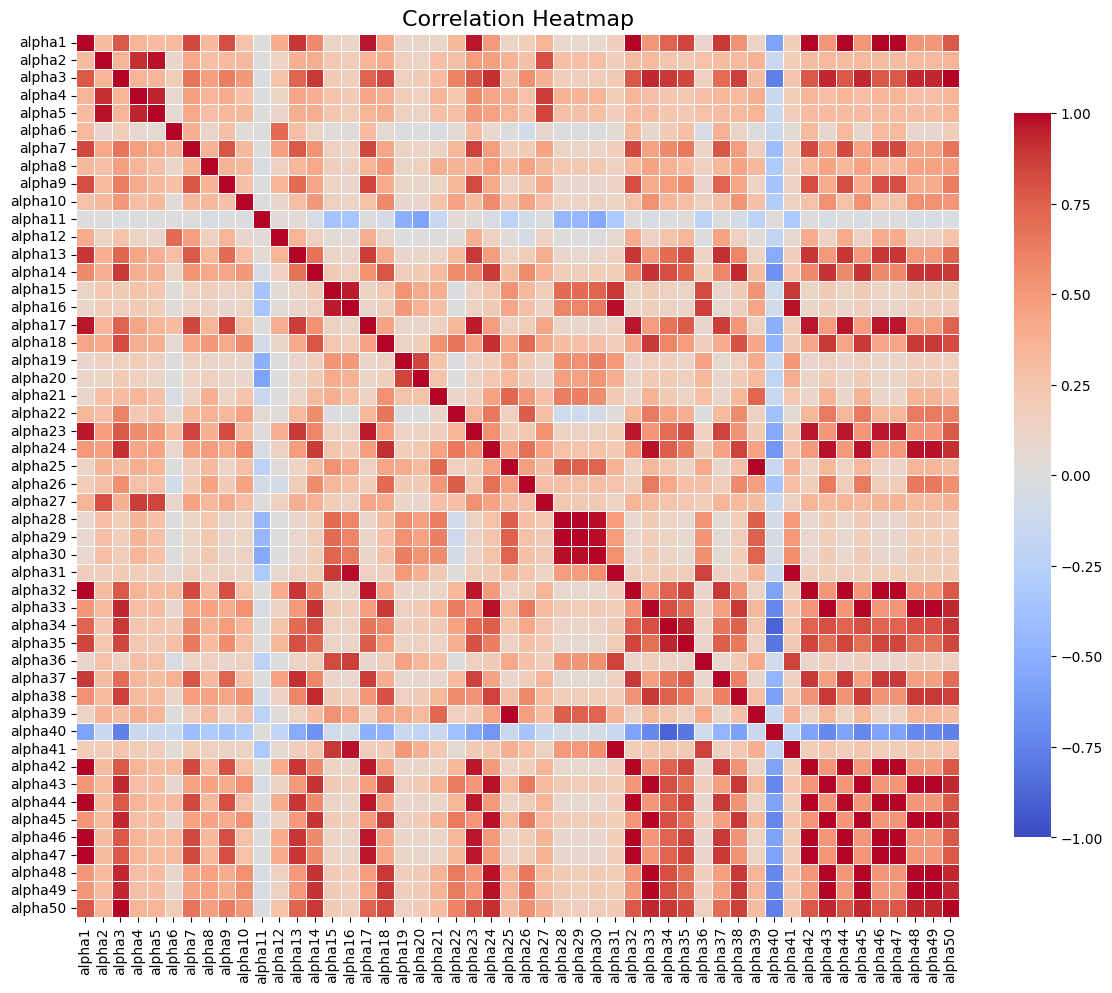}
    \caption{Toyota}
  \end{subfigure}
    \hfill
  \begin{subfigure}{0.45\textwidth}
    \centering
    \includegraphics[width=\linewidth]{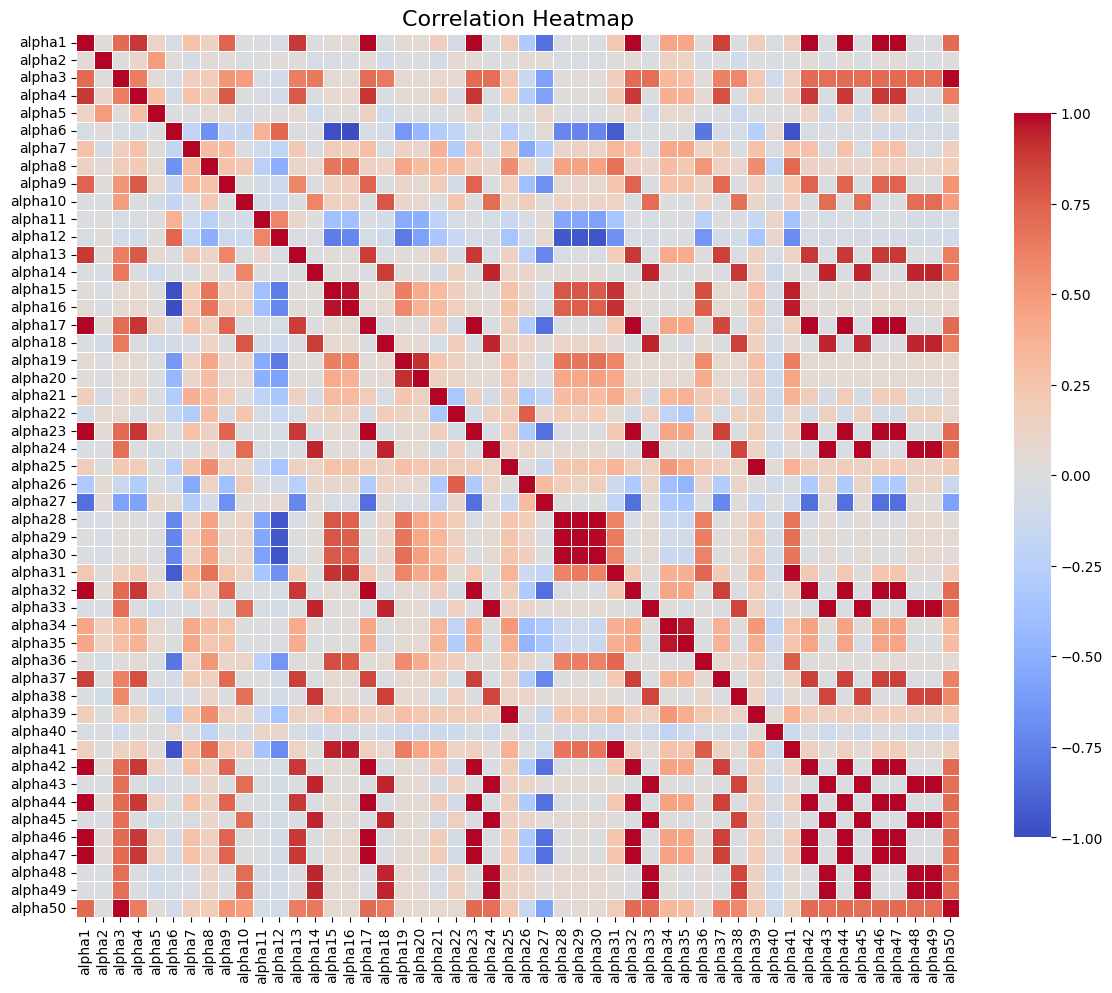}
    \caption{Pepsi}
  \end{subfigure}
  
  \caption{Correlation Heatmaps of 50 LLM-generated Alphas}
  \label{fig:corr_heatmap}
\end{figure*}

\begin{figure*}[!htbp]
  \centering
  % First row
  \begin{subfigure}{0.45\textwidth}
    \centering
    \includegraphics[width=\linewidth]{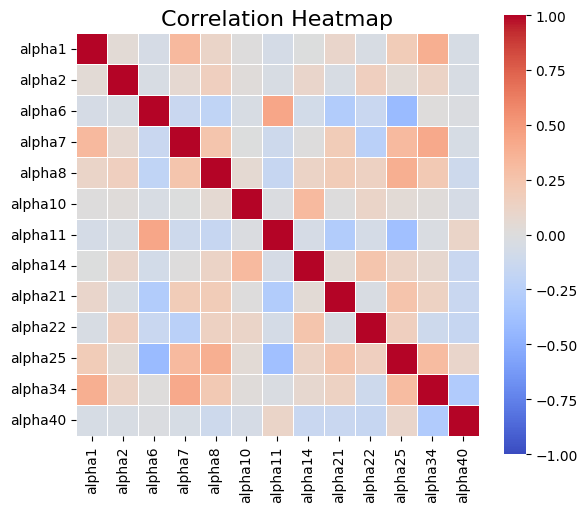}
    \caption{Apple}
  \end{subfigure}
  \hfill
  \begin{subfigure}{0.45\textwidth}
    \centering
    \includegraphics[width=\linewidth]{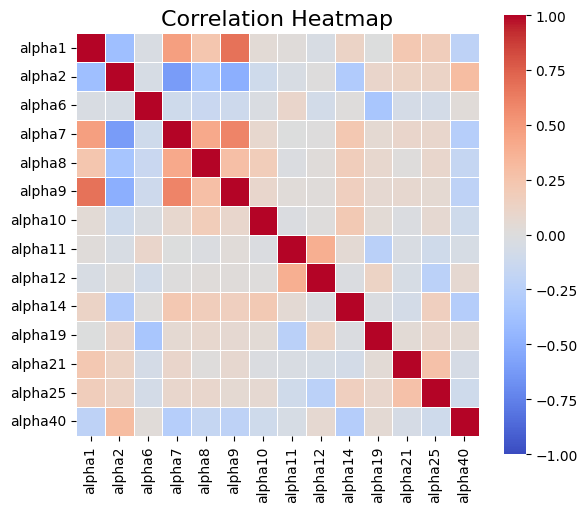}
    \caption{HSBC}
  \end{subfigure}
  \hfill
  \begin{subfigure}{0.45\textwidth}
    \centering
    \includegraphics[width=\linewidth]{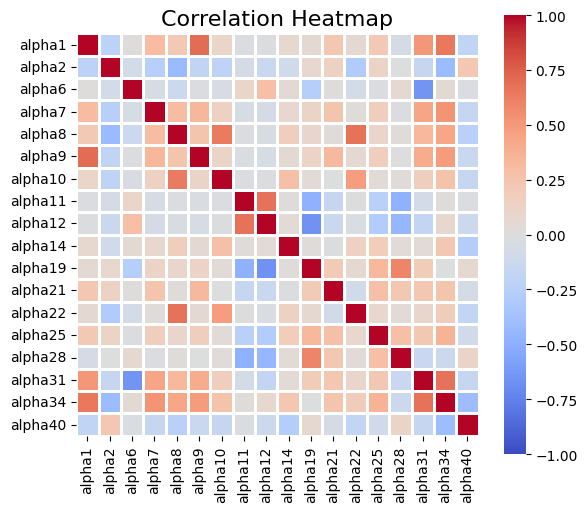}
    \caption{Tencent}
  \end{subfigure}
  \hfill
  \begin{subfigure}{0.45\textwidth}
    \centering
    \includegraphics[width=\linewidth]{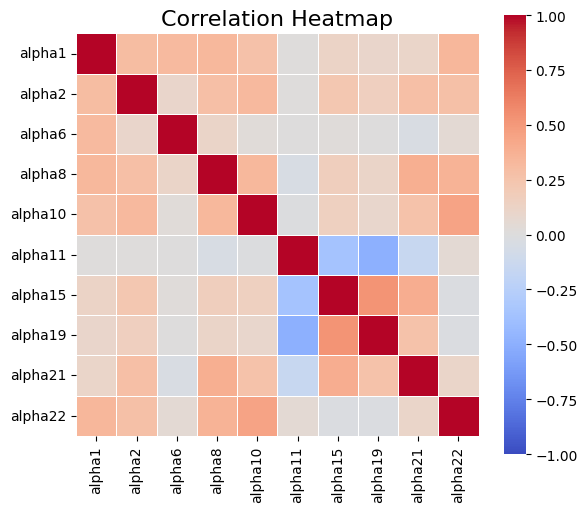}
    \caption{Toyota}
  \end{subfigure}
    \hfill
  \begin{subfigure}{0.45\textwidth}
    \centering
    \includegraphics[width=\linewidth]{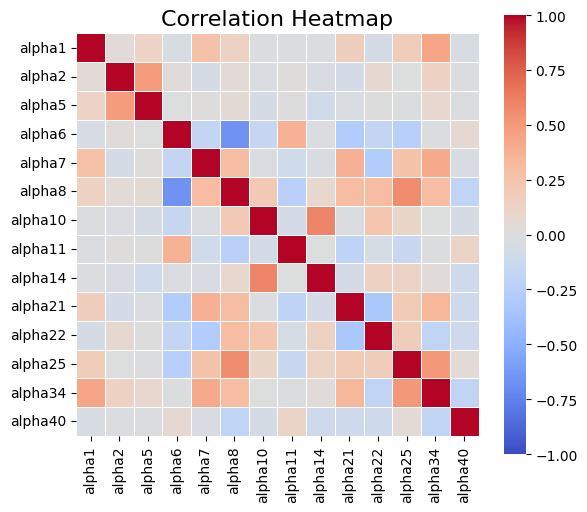}
    \caption{Pepsi}
  \end{subfigure}
  
  \caption{Correlation Heatmaps of LLM-generated Alphas ($\left| \text{Corr} \right|$ $\leq 0.7$)}
  \label{fig:corr_heatmap corr <= 0.03}
\end{figure*}

To evaluate how the number of formulaic alphas influences trading performance, we conduct a more detailed analysis by adjusting the number of LLM-generated alphas fed into the reinforcement learning module. Three different methods are used. For the first approach, we reduce redundancy by measuring the pairwise correlations between them. Highly correlated alphas often carry overlapping information, which may not contribute much to the learning process and could even introduce noise. To address this, we remove alphas with the absolute value of pairwise correlation coefficients (Figure~\ref{fig:corr_heatmap}) exceeding a threshold of 0.7 (Figure~\ref{fig:corr_heatmap corr <= 0.03}), a common choice in quantitative research. This helps retain only the most diverse and informative signals for the learning process. For the second approach, we choose the top ten alphas with the highest contribution in terms of the LightGBM feature importance (Gain) (Figure~\ref{fig:FI}). For the third approach, thirty alphas are chosen randomly for each company.

Table~\ref{tab:different_alpha_selection_method} presents the results when using different alpha selection methods. First of all, when comparing the PPO strategy using all 50 alphas with the version that applies low-correlation filtering, we observe that reducing redundancy among alphas affects stocks differently. For Apple and Tencent, performance declines slightly in terms of returns or risk-adjusted metrics, suggesting that some removed alphas may have contributed useful information. In contrast, HSBC, Pepsi, and Toyota show modest improvements, indicating that the elimination of highly correlated alphas can slightly enhance the signal quality for these stocks. In general, the low-correlation approach maintains the ability of the strategy to outperform market benchmarks while slightly increasing variability for certain stocks, reflecting a trade-off between signal diversity and stability.

When it comes to performance using high-contribution alphas, Apple, Pepsi, and Tencent experience slight declines in cumulative returns and Sharpe ratios, along with  higher drawdowns, suggesting that focusing solely on the top-contributing alphas may reduce the diversity of useful signals for these stocks. In contrast, HSBC and Toyota see improvements in both return and risk-adjusted performance, indicating that high-contribution alphas provide particularly strong signals for these assets. Overall, while the high-contribution approach maintains the ability to outperform market benchmarks, it does not consistently improve performance across all stocks and may sacrifice stability for certain assets.

Furthermore, when using thirty randomly selected alphas, Apple continues to exhibit strong performance, achieving the highest cumulative return and Sharpe ratio among all five stocks. In contrast, Toyota still performs the worst in this setting. Notably, HSBC shows improved performance compared to the scenario using all fifty alphas in terms of return and Sharpe ratio. 

In general, although the impact varies across individual stocks, the different alpha selection settings do not substantially alter the overall results.

\subsection{Impact of Prompt Information on Trading Performance} \label{prompt information}

In this subsection, we examine how formulaic alphas generated with different prompt information influence the performance of the proposed RL strategy. Originally, the prompt is fed a pandas DataFrame in JSON format containing various features of all companies. Here, we consider two additional scenarios: (1) providing the prompt with only feature names and no additional data, instructing the model to generate alphas based solely on its own knowledge; and (2) providing the prompt with partial information, such as data from a single company.

For simplicity, the analysis focuses solely on Apple. In the partial information scenario, only data from Apple is provided. Table~\ref{tab:different_prompt_setting} demonstrates the PPO strategy performance under different scenarios. Based on the results, we observe that across all scenarios, formulaic alphas generated with varying prompt information produce strong performance under the proposed RL framework, with Sharpe ratios above one, demonstrating the robustness and effectiveness of the framework.

\begin{table}[htbp]
    \centering
    \caption{Performance of PPO-Adjusted Strategy on Apple under Different Prompt Information. Mean (Standard Deviation).}
    \label{tab:different_prompt_setting}
        \begin{tabular}{|l|c|c|c|}
            \hline
            \textbf{Metric} & \textbf{Full Information} & \textbf{Partial Information} & \textbf{Feature Name Only} \\ \hline
            Cumulative Return & 1.6817 (0.0619) & 1.4312 (0.1342) & 1.2134 (0.0848) \\ \hline
            Sharpe Ratio & 1.9998 (0.0169) & 1.9715 (0.0622) & 1.9540 (0.0360) \\ \hline
            Drawdown & $-0.0101$ (0.0007) & $-0.0130$ (0.0011) & $-0.0116$ (0.0012) \\ \hline
        \end{tabular}
\end{table}

\subsection{Impact of Sentiment on Trading Performance}

To evaluate the impact of sentiment on trading performance, we consider three settings: (1) sentiment of the target company only, (2) sentiment of all companies, and (3) no sentiment. For each target company, fifty unique alphas are generated in each scenario. As demonstrated in Section~\ref{prompt information}, using feature names alone is sufficient to achieve strong trading performance. Therefore, in this section, we simplify the prompts by including only feature names.

Table~\ref{tab:combined_sentiment_analysis} reports the performance of the proposed PPO strategy for all five companies under different sentiment settings. The results suggest that incorporating sentiment features within the PPO framework does not lead to substantial differences in trading performance. Across all three sentiment settings for each target company, the proposed strategy consistently delivers strong results.

\begin{table}[htbp]
\centering
\small
\caption{Performance Comparison of PPO-Adjusted Strategy across Different Sentiment Settings. Values are Mean (Standard Deviation).}
\label{tab:combined_sentiment_analysis}

\begin{tabular}{l l ccc}
\toprule
\textbf{Stock} & \textbf{Metric} 
& \textbf{Target Only} 
& \textbf{All Companies} 
& \textbf{No Sentiment} \\
\midrule

\multirow{3}{*}{Apple}
& Cumulative Return & 1.2499 (0.8023) & 1.4797 (0.1156) & 1.4459 (0.1098) \\
& Sharpe Ratio      & 1.9676 (0.0344) & 2.0134 (0.0421) & 1.9933 (0.0387) \\
& Max Drawdown      & $-0.0133$ (0.0013) & $-0.0084$ (0.0019) & $-0.0110$ (0.0020) \\
\midrule

\multirow{3}{*}{HSBC}
& Cumulative Return & 0.6093 (0.0455) & 0.4460 (0.0525) & 0.5955 (0.0481) \\
& Sharpe Ratio      & 1.2690 (0.2100) & 0.9343 (0.1959) & 1.1638 (0.2021) \\
& Max Drawdown      & $-0.1014$ (0.0239) & $-0.1597$ (0.0311) & $-0.1489$ (0.0283) \\
\midrule

\multirow{3}{*}{Pepsi}
& Cumulative Return & 0.4124 (0.0396) & 0.7643 (0.0514) & 0.7644 (0.0437) \\
& Sharpe Ratio      & 1.7663 (0.1956) & 1.7786 (0.1822) & 1.7972 (0.1891) \\
& Max Drawdown      & $-0.0454$ (0.0111) & $-0.0051$ (0.0070) & $-0.0047$ (0.0064) \\
\midrule

\multirow{3}{*}{Tencent}
& Cumulative Return & 0.6136 (0.0486) & 0.7163 (0.0436) & 0.7067 (0.0513) \\
& Sharpe Ratio      & 1.1961 (0.1821) & 1.3742 (0.1953) & 1.2433 (0.1779) \\
& Max Drawdown      & $-0.0515$ (0.0120) & $-0.0504$ (0.0132) & $-0.0484$ (0.0112) \\
\midrule

\multirow{3}{*}{Toyota}
& Cumulative Return & 0.0272 (0.0554) & 0.1088 (0.0622) & 0.0102 (0.0512) \\
& Sharpe Ratio      & 0.0982 (0.0326) & 0.2549 (0.0451) & 0.0641 (0.0287) \\
& Max Drawdown      & $-0.2735$ (0.0413) & $-0.2203$ (0.0370) & $-0.2651$ (0.0395) \\
\bottomrule
\end{tabular}

\end{table}

%--- Section ---%
\section{Conclusion}\label{sec5}

This study explores the use of large language models to generate formulaic alphas and applies PPO to dynamically optimize their weights for trading decisions. Empirical results show that, while the PPO-adjusted strategy does not consistently achieve the highest cumulative returns compared with traditional benchmarks such as buy-and-hold and momentum strategies, it demonstrates clear advantages in terms of risk-adjusted performance. Specifically, the PPO strategy achieves higher Sharpe ratios across most target stocks, indicating more stable return profiles relative to volatility. In addition, the strategy consistently exhibits lower maximum drawdowns, suggesting improved downside risk control and enhanced robustness during adverse market conditions. Compared with an equal-weighted alpha portfolio, the PPO-based approach further highlights the benefit of adaptive weight allocation in balancing return and risk. Overall, these findings suggest that integrating LLM-generated alphas with reinforcement learning offers a promising framework for constructing more risk-aware and resilient trading strategies, particularly in settings where drawdown control and return stability are prioritized over absolute returns.

Furthermore, the ablation analysis further clarifies the roles of both alpha construction and reinforcement learning algorithms within the proposed framework. For most of the selected stocks, portfolios built on LLM-generated alphas consistently exhibit superior performance compared to those based on human-crafted alphas. In terms of optimization methods, alternative reinforcement learning algorithms, including SAC, TD3, and A2C, achieve performance comparable to PPO under the same environmental setup.

Besides, reducing the number of alphas leads to slight improvements in both cumulative return and Sharpe ratio for certain stocks. However, for others, their performance may worsen after reducing the number of alphas. This indicates that selecting a well-balanced set of informative and diverse alphas can enhance both stability and profitability. The results underscore the potential of combining LLMs with reinforcement learning to create effective and interpretable trading strategies. 

This study has several limitations. First, the analysis is conducted on only ten stocks, which may limit the generalizability of the findings. Future work will further extend the analysis to a larger and more diverse set of stocks. Second, due to limited access to high-frequency market data, the current experiments are based on daily observations, which constrains the practical applicability of the proposed LLM-based framework in real-world trading environments where higher-frequency information and richer factor signals are often available. Future studies will consider high-frequency data and incorporate a broader range of explanatory factors, including macroeconomic variables and firm-level characteristics, to better capture market dynamics. In addition, future work will explore alternative LLM architectures, and broader applications across different markets and asset classes.

%Bibliography
\bibliographystyle{unsrt}  
\bibliography{references}

@INPROCEEDINGS{9580172,
  author={Pardeshi, Yash K. and Kale, Prof. Preeti},
  booktitle={2021 12th International Conference on Computing Communication and Networking Technologies (ICCCNT)}, 
  title={Technical Analysis Indicators in Stock Market Using Machine Learning: A Comparative Analysis}, 
  year={2021},
  volume={},
  number={},
  pages={1-6},
  doi={10.1109/ICCCNT51525.2021.9580172}}

@article{MOSTAFAVI2025100631,
title = {Key technical indicators for stock market prediction},
journal = {Machine Learning with Applications},
volume = {20},
pages = {100631},
year = {2025},
issn = {2666-8270},
doi = {https://doi.org/10.1016/j.mlwa.2025.100631},
url = {https://www.sciencedirect.com/science/article/pii/S2666827025000143},
author = {Seyed Mostafa Mostafavi and Ali Reza Hooman}
}

@article{MINTARYA202396,
title = {Machine learning approaches in stock market prediction: A systematic literature review},
journal = {Procedia Computer Science},
volume = {216},
pages = {96-102},
year = {2023},
note = {7th International Conference on Computer Science and Computational Intelligence 2022},
issn = {1877-0509},
doi = {https://doi.org/10.1016/j.procs.2022.12.115},
url = {https://www.sciencedirect.com/science/article/pii/S1877050922021937},
author = {Latrisha N. Mintarya and Jeta N.M. Halim and Callista Angie and Said Achmad and Aditya Kurniawan}
}

@ARTICLE{10114634,
  author={Gao, Lei and Guan, Ling},
  journal={IEEE MultiMedia}, 
  title={Interpretability of Machine Learning: Recent Advances and Future Prospects}, 
  year={2023},
  volume={30},
  number={4},
  pages={105-118},
  doi={10.1109/MMUL.2023.3272513}}

@misc{kou2024automatestrategyfindingllm,
      title={Automate Strategy Finding with LLM in Quant investment}, 
      author={Zhizhuo Kou and Holam Yu and Jingshu Peng and Lei Chen},
      year={2024},
      eprint={2409.06289},
      archivePrefix={arXiv},
      primaryClass={q-fin.PM},
      url={https://arxiv.org/abs/2409.06289}, 
}

@INPROCEEDINGS{10825946,
  author={Chen, Qizhao and Kawashima, Hiroaki},
  booktitle={2024 IEEE International Conference on Big Data (BigData)}, 
  title={Stock Price Prediction Using LLM-Based Sentiment Analysis}, 
  year={2024},
  volume={},
  number={},
  pages={4846-4853},
  doi={10.1109/BigData62323.2024.10825946}}

@article{FAMA19933,
title = {Common risk factors in the returns on stocks and bonds},
journal = {Journal of Financial Economics},
volume = {33},
number = {1},
pages = {3-56},
year = {1993},
issn = {0304-405X},
doi = {https://doi.org/10.1016/0304-405X(93)90023-5},
url = {https://www.sciencedirect.com/science/article/pii/0304405X93900235},
author = {Eugene F. Fama and Kenneth R. French}
}

@INPROCEEDINGS{8250694,
  author={Kamble, Rupesh A.},
  booktitle={2017 International Conference on Intelligent Computing and Control Systems (ICICCS)}, 
  title={Short and long term stock trend prediction using decision tree}, 
  year={2017},
  volume={},
  number={},
  pages={1371-1375},
  doi={10.1109/ICCONS.2017.8250694}}

@INPROCEEDINGS{6703096,
  author={Hu, Zhen and Zhu, Jie and Tse, Ken},
  booktitle={2013 6th International Conference on Information Management, Innovation Management and Industrial Engineering}, 
  title={Stocks market prediction using Support Vector Machine}, 
  year={2013},
  volume={2},
  number={},
  pages={115-118},
  doi={10.1109/ICIII.2013.6703096}}

@INPROCEEDINGS{9361804,
  author={Liu, Zixuan and Dang, Ziyuan and Yu, Jie},
  booktitle={2020 International Conference on Computer Engineering and Intelligent Control (ICCEIC)}, 
  title={Stock Price Prediction Model Based on RBF-SVM Algorithm}, 
  year={2020},
  volume={},
  number={},
  pages={124-127},
  doi={10.1109/ICCEIC51584.2020.00032}}

@INPROCEEDINGS{9987903,
  author={Du, Shipei and Hao, Dehong and Li, Xiao},
  booktitle={2022 IEEE 2nd International Conference on Data Science and Computer Application (ICDSCA)}, 
  title={Research on stock forecasting based on random forest}, 
  year={2022},
  volume={},
  number={},
  pages={301-305},
  doi={10.1109/ICDSCA56264.2022.9987903}}

@INPROCEEDINGS{9317207,
  author={Mehtab, Sidra and Sen, Jaydip},
  booktitle={2020 International Conference on Decision Aid Sciences and Application (DASA)}, 
  title={Stock Price Prediction Using CNN and LSTM-Based Deep Learning Models}, 
  year={2020},
  volume={},
  number={},
  pages={447-453},
  doi={10.1109/DASA51403.2020.9317207}}

@INPROCEEDINGS{10392023,
  author={Shinde, Sagar and Wadhwa, Lalitkumar and Mohane, Naynesh and Pagar, Vishal and Sherje, Nitin and Mane, Sohan},
  booktitle={2023 7th International Conference On Computing, Communication, Control And Automation (ICCUBEA)}, 
  title={Stock Price Prediction using LSTM}, 
  year={2023},
  volume={},
  number={},
  pages={1-7},
  doi={10.1109/ICCUBEA58933.2023.10392023}}

@Article{math11091985,
AUTHOR = {Zhang, Jilin and Ye, Lishi and Lai, Yongzeng},
TITLE = {Stock Price Prediction Using CNN-BiLSTM-Attention Model},
JOURNAL = {Mathematics},
VOLUME = {11},
YEAR = {2023},
NUMBER = {9},
ARTICLE-NUMBER = {1985},
URL = {https://www.mdpi.com/2227-7390/11/9/1985},
ISSN = {2227-7390},
DOI = {10.3390/math11091985}
}

@article{https://doi.org/10.1155/2020/6622927,
author = {Lu, Wenjie and Li, Jiazheng and Li, Yifan and Sun, Aijun and Wang, Jingyang},
title = {A CNN-LSTM-Based Model to Forecast Stock Prices},
journal = {Complexity},
volume = {2020},
number = {1},
pages = {6622927},
doi = {https://doi.org/10.1155/2020/6622927},
year = {2020}
}

@misc{zhou2021informerefficienttransformerlong,
      title={Informer: Beyond Efficient Transformer for Long Sequence Time-Series Forecasting}, 
      author={Haoyi Zhou and Shanghang Zhang and Jieqi Peng and Shuai Zhang and Jianxin Li and Hui Xiong and Wancai Zhang},
      year={2021},
      eprint={2012.07436},
      archivePrefix={arXiv},
      primaryClass={cs.LG},
      url={https://arxiv.org/abs/2012.07436}, 
}

@misc{wu2022autoformerdecompositiontransformersautocorrelation,
      title={Autoformer: Decomposition Transformers with Auto-Correlation for Long-Term Series Forecasting}, 
      author={Haixu Wu and Jiehui Xu and Jianmin Wang and Mingsheng Long},
      year={2022},
      eprint={2106.13008},
      archivePrefix={arXiv},
      primaryClass={cs.LG},
      url={https://arxiv.org/abs/2106.13008}, 
}

@misc{wang2024timexerempoweringtransformerstime,
      title={TimeXer: Empowering Transformers for Time Series Forecasting with Exogenous Variables}, 
      author={Yuxuan Wang and Haixu Wu and Jiaxiang Dong and Guo Qin and Haoran Zhang and Yong Liu and Yunzhong Qiu and Jianmin Wang and Mingsheng Long},
      year={2024},
      eprint={2402.19072},
      archivePrefix={arXiv},
      primaryClass={cs.LG},
      url={https://arxiv.org/abs/2402.19072}, 
}

@misc{li2023mastermarketguidedstocktransformer,
      title={MASTER: Market-Guided Stock Transformer for Stock Price Forecasting}, 
      author={Tong Li and Zhaoyang Liu and Yanyan Shen and Xue Wang and Haokun Chen and Sen Huang},
      year={2023},
      eprint={2312.15235},
      archivePrefix={arXiv},
      primaryClass={cs.CE},
      url={https://arxiv.org/abs/2312.15235}, 
}

@misc{yang2023fingptopensourcefinanciallarge,
      title={FinGPT: Open-Source Financial Large Language Models}, 
      author={Hongyang Yang and Xiao-Yang Liu and Christina Dan Wang},
      year={2023},
      eprint={2306.06031},
      archivePrefix={arXiv},
      primaryClass={q-fin.ST},
      url={https://arxiv.org/abs/2306.06031}, 
}

@misc{xie2023pixiulargelanguagemodel,
      title={PIXIU: A Large Language Model, Instruction Data and Evaluation Benchmark for Finance}, 
      author={Qianqian Xie and Weiguang Han and Xiao Zhang and Yanzhao Lai and Min Peng and Alejandro Lopez-Lira and Jimin Huang},
      year={2023},
      eprint={2306.05443},
      archivePrefix={arXiv},
      primaryClass={cs.CL},
      url={https://arxiv.org/abs/2306.05443}, 
}

@misc{kou2025automatestrategyfindingllm,
      title={Automate Strategy Finding with LLM in Quant Investment}, 
      author={Zhizhuo Kou and Holam Yu and Junyu Luo and Jingshu Peng and Lei Chen},
      year={2025},
      eprint={2409.06289},
      archivePrefix={arXiv},
      primaryClass={q-fin.PM},
      url={https://arxiv.org/abs/2409.06289}, 
}

@misc{wang2024gptsignalgenerativeaisemiautomated,
      title={GPT-Signal: Generative AI for Semi-automated Feature Engineering in the Alpha Research Process}, 
      author={Yining Wang and Jinman Zhao and Yuri Lawryshyn},
      year={2024},
      eprint={2410.18448},
      archivePrefix={arXiv},
      primaryClass={cs.CE},
      url={https://arxiv.org/abs/2410.18448}, 
}

@misc{schulman2017proximalpolicyoptimizationalgorithms,
      title={Proximal Policy Optimization Algorithms}, 
      author={John Schulman and Filip Wolski and Prafulla Dhariwal and Alec Radford and Oleg Klimov},
      year={2017},
      eprint={1707.06347},
      archivePrefix={arXiv},
      primaryClass={cs.LG},
      url={https://arxiv.org/abs/1707.06347}, 
}

@misc{kakushadze2016101formulaicalphas,
      title={101 Formulaic Alphas}, 
      author={Zura Kakushadze},
      year={2016},
      eprint={1601.00991},
      archivePrefix={arXiv},
      primaryClass={q-fin.PM},
      url={https://arxiv.org/abs/1601.00991}, 
}

@article{Sejnowski_2020,
   title={The unreasonable effectiveness of deep learning in artificial intelligence},
   volume={117},
   ISSN={1091-6490},
   url={http://dx.doi.org/10.1073/pnas.1907373117},
   DOI={10.1073/pnas.1907373117},
   number={48},
   journal={Proceedings of the National Academy of Sciences},
   publisher={Proceedings of the National Academy of Sciences},
   author={Sejnowski, Terrence J.},
   year={2020},
   month=jan, pages={30033–30038} }

@INPROCEEDINGS{8279188,
  author={Zhou, Yi and Lin, Jianwu},
  booktitle={2017 IEEE/SICE International Symposium on System Integration (SII)}, 
  title={The alpha life cycle of quantitative strategy}, 
  year={2017},
  volume={},
  number={},
  pages={53-59},
  doi={10.1109/SII.2017.8279188}}

@inproceedings{Chen2025,
  author    = {Chen, Qizhao},
  title     = {Stock Price Change Prediction Using Prompt-Based {LLM}s with {RL}-Enhanced Post-Hoc Adjustments},
  booktitle = {Advances in Intelligent Systems Research},
  pages     = {475--483},
  publisher = {Atlantis Press International BV},
  year      = {2025},
  doi       = {10.2991/978-94-6463-742-7_46},
  address = {Paris}
}

@misc{huang2025deepreinforcementlearningframework,
      title={A Deep Reinforcement Learning Framework for Dynamic Portfolio Optimization: Evidence from China's Stock Market}, 
      author={Gang Huang and Xiaohua Zhou and Qingyang Song},
      year={2025},
      eprint={2412.18563},
      archivePrefix={arXiv},
      primaryClass={q-fin.PM},
      url={https://arxiv.org/abs/2412.18563}, 
}

@misc{ndikum2024advancinginvestmentfrontiersindustrygrade,
      title={Advancing Investment Frontiers: Industry-grade Deep Reinforcement Learning for Portfolio Optimization}, 
      author={Philip Ndikum and Serge Ndikum},
      year={2024},
      eprint={2403.07916},
      archivePrefix={arXiv},
      primaryClass={cs.AI},
      url={https://arxiv.org/abs/2403.07916}, 
}

@INPROCEEDINGS{10627674,
  author={Sharma, Rishabh and Sharma, Ajay and Hariharan, Shanmugasundaram and Jain, Vishal},
  booktitle={2024 4th International Conference on Intelligent Technologies (CONIT)}, 
  title={Adaptive Investment Strategies: Deep Reinforcement Learning Approaches for Portfolio Optimization}, 
  year={2024},
  volume={},
  number={},
  pages={1-5},
  doi={10.1109/CONIT61985.2024.10627674}}

@Article{ijfs11010010,
AUTHOR = {Millea, Adrian and Edalat, Abbas},
TITLE = {Using Deep Reinforcement Learning with Hierarchical Risk Parity for Portfolio Optimization},
JOURNAL = {International Journal of Financial Studies},
VOLUME = {11},
YEAR = {2023},
NUMBER = {1},
ARTICLE-NUMBER = {10},
URL = {https://www.mdpi.com/2227-7072/11/1/10},
ISSN = {2227-7072},
DOI = {10.3390/ijfs11010010}
}

@misc{hinton2015distillingknowledgeneuralnetwork,
      title={Distilling the Knowledge in a Neural Network}, 
      author={Geoffrey Hinton and Oriol Vinyals and Jeff Dean},
      year={2015},
      eprint={1503.02531},
      archivePrefix={arXiv},
      primaryClass={stat.ML},
      url={https://arxiv.org/abs/1503.02531}, 
}

@book{dreman1998contrarian,
  title={Contrarian Investment Strategies: The Next Generation},
  author={David Dreman},
  year={1998},
  publisher={Simon \& Schuster},
  address={New York}
}

@article{chen2025sentiment,
  author  = {Chen, Qizhao and Kawashima, Hiroshi},
  title   = {A novel sentiment correlation-based method with dual transformer model for stock price prediction},
  journal = {International Journal of Data Science and Analytics},
  volume  = {21},
  number  = {1},
  year    = {2025},
  doi     = {10.1007/s41060-025-00932-7}
}

@misc{chen2025sentimentawarestockpriceprediction,
      title={Sentiment-Aware Stock Price Prediction with Transformer and LLM-Generated Formulaic Alpha}, 
      author={Qizhao Chen and Hiroaki Kawashima},
      year={2025},
      eprint={2508.04975},
      archivePrefix={arXiv},
      primaryClass={cs.CE},
      url={https://arxiv.org/abs/2508.04975}, 
}

@inproceedings{10.5555/3491440.3492067,
author = {Lin, Siyu and Beling, Peter A.},
title = {An end-to-end optimal trade execution framework based on proximal policy optimization},
year = {2021},
isbn = {9780999241165},
booktitle = {Proceedings of the Twenty-Ninth International Joint Conference on Artificial Intelligence},
articleno = {627},
numpages = {7},
location = {Yokohama, Yokohama, Japan},
series = {IJCAI'20}
}

@ARTICLE{10703056,
  author={Rio, Alberto del and Jimenez, David and Serrano, Javier},
  journal={IEEE Access}, 
  title={Comparative Analysis of A3C and PPO Algorithms in Reinforcement Learning: A Survey on General Environments}, 
  year={2024},
  volume={12},
  number={},
  pages={146795-146806},
  doi={10.1109/ACCESS.2024.3472473}}

@article{ZHANG2022750,
title = {Proximal policy optimization via enhanced exploration efficiency},
journal = {Information Sciences},
volume = {609},
pages = {750-765},
year = {2022},
issn = {0020-0255},
doi = {https://doi.org/10.1016/j.ins.2022.07.111},
url = {https://www.sciencedirect.com/science/article/pii/S0020025522008003},
author = {Junwei Zhang and Zhenghao Zhang and Shuai Han and Shuai Lyu}
}

@article{Mammadzada2025PPO,
  author  = {Mammadzada, R.},
  title   = {Reinforcement Learning in a Virtual World: A Study of PPO and SAC within Unity ML-Agents},
  journal = {Chemical Technology, Control and Management},
  year    = {2025},
  number  = {5},
  pages   = {96--103},
  doi     = {10.59048/2181-1105.1722}
}

@inproceedings{chen2025llmstock,
  author    = {Chen, Qizhao},
  title     = {Stock Price Prediction with {LLM}-Guided Market Movement Signals and Transformer Model},
  booktitle = {FinTech and Sustainable Innovation},
  year      = {2025},
  doi       = {10.47852/bonviewfsi52025703}
}

@article{chen2025image,
  author  = {Chen, Qizhao},
  title   = {Image-Driven Stock Price Prediction with {LLaMA}: A Prompt-Based Approach},
  journal = {International Journal of Modeling and Optimization},
  pages   = {17--24},
  year    = {2025},
  doi     = {10.7763/ijmo.2025.v15.867}
}

@article{chen2025anomaly,
  author  = {Chen, Qizhao},
  title   = {Explore Anomaly-Aware Transformers for Robust Financial Time Series Forecasting},
  journal = {Journal of Computer and Communications},
  volume  = {13},
  number  = {12},
  pages   = {100--114},
  year    = {2025},
  doi     = {10.4236/jcc.2025.1312006}
}

@misc{chen2025frameworkmeasuringnewstopics,
      title={A Framework for Measuring How News Topics Drive Stock Movement}, 
      author={Qizhao Chen},
      year={2025},
      eprint={2510.06864},
      archivePrefix={arXiv},
      primaryClass={cs.CE},
      url={https://arxiv.org/abs/2510.06864}, 
}

@ARTICLE{10285085,
  author={Lee, Namyeong and Moon, Jun},
  journal={IEEE Access}, 
  title={Offline Reinforcement Learning for Automated Stock Trading}, 
  year={2023},
  volume={11},
  number={},
  pages={112577-112589},
  doi={10.1109/ACCESS.2023.3324458}}

@article{orra2025reward,
  author  = {Orra, A. and Choudhary, H. and Sharma, A. and Thakur, M.},
  title   = {Enhancing Deep Reinforcement Learning for Stock Trading: A Reward Shaping Approach via Expert Feedback},
  journal = {Knowledge and Information Systems},
  volume  = {67},
  number  = {11},
  pages   = {11075--11094},
  year    = {2025},
  doi     = {10.1007/s10115-025-02562-8}
}

@Article{app13010633,
AUTHOR = {Kong, Minseok and So, Jungmin},
TITLE = {Empirical Analysis of Automated Stock Trading Using Deep Reinforcement Learning},
JOURNAL = {Applied Sciences},
VOLUME = {13},
YEAR = {2023},
NUMBER = {1},
ARTICLE-NUMBER = {633},
URL = {https://www.mdpi.com/2076-3417/13/1/633},
ISSN = {2076-3417},
DOI = {10.3390/app13010633}
}

@article{guevara2025actorcritic,
  author  = {Guevara, C.},
  title   = {Stock Market Trading via Actor--Critic Reinforcement Learning and Adaptable Data Structure},
  journal = {PeerJ Computer Science},
  volume  = {11},
  pages   = {e2690},
  year    = {2025},
  doi     = {10.7717/peerj-cs.2690}
}

@ARTICLE{10904473,
  author={Sattar, Asma and Sarwar, Amna and Gillani, Saira and Bukhari, Maryam and Rho, Seungmin and Faseeh, Muhammad},
  journal={IEEE Access}, 
  title={A Novel RMS-Driven Deep Reinforcement Learning for Optimized Portfolio Management in Stock Trading}, 
  year={2025},
  volume={13},
  number={},
  pages={42813-42835},
  doi={10.1109/ACCESS.2025.3546099}}

@article{VIJH2020599,
title = {Stock Closing Price Prediction using Machine Learning Techniques},
journal = {Procedia Computer Science},
volume = {167},
pages = {599-606},
year = {2020},
note = {International Conference on Computational Intelligence and Data Science},
issn = {1877-0509},
doi = {https://doi.org/10.1016/j.procs.2020.03.326},
url = {https://www.sciencedirect.com/science/article/pii/S1877050920307924},
author = {Mehar Vijh and Deeksha Chandola and Vinay Anand Tikkiwal and Arun Kumar}
}

@Article{electronics9091384,
AUTHOR = {Yuan, Yuyu and Wen, Wen and Yang, Jincui},
TITLE = {Using Data Augmentation Based Reinforcement Learning for Daily Stock Trading},
JOURNAL = {Electronics},
VOLUME = {9},
YEAR = {2020},
NUMBER = {9},
ARTICLE-NUMBER = {1384},
URL = {https://www.mdpi.com/2079-9292/9/9/1384},
ISSN = {2079-9292},
DOI = {10.3390/electronics9091384}
}

@ARTICLE{10195852,
  author={Fu, Kui and Yu, Yidong and Li, Bing},
  journal={IEEE Access}, 
  title={Multi-Feature Supervised Reinforcement Learning for Stock Trading}, 
  year={2023},
  volume={11},
  number={},
  pages={77840-77855},
  doi={10.1109/ACCESS.2023.3298821}}

@inproceedings{10.1145/3529836.3529857,
author = {GE, Jun and QIN, Yuanqi and Li, Yaling and Huang, yanjia and Hu, Hao},
title = {Single stock trading with deep reinforcement learning: A comparative study},
year = {2022},
isbn = {9781450395700},
publisher = {Association for Computing Machinery},
address = {New York, NY, USA},
url = {https://doi.org/10.1145/3529836.3529857},
doi = {10.1145/3529836.3529857},
booktitle = {Proceedings of the 2022 14th International Conference on Machine Learning and Computing},
pages = {34–43},
numpages = {10},
location = {Guangzhou, China},
series = {ICMLC '22}
}

@misc{mnih2016asynchronousmethodsdeepreinforcement,
      title={Asynchronous Methods for Deep Reinforcement Learning}, 
      author={Volodymyr Mnih and Adrià Puigdomènech Badia and Mehdi Mirza and Alex Graves and Timothy P. Lillicrap and Tim Harley and David Silver and Koray Kavukcuoglu},
      year={2016},
      eprint={1602.01783},
      archivePrefix={arXiv},
      primaryClass={cs.LG},
      url={https://arxiv.org/abs/1602.01783}, 
}

@misc{haarnoja2018softactorcriticoffpolicymaximum,
      title={Soft Actor-Critic: Off-Policy Maximum Entropy Deep Reinforcement Learning with a Stochastic Actor}, 
      author={Tuomas Haarnoja and Aurick Zhou and Pieter Abbeel and Sergey Levine},
      year={2018},
      eprint={1801.01290},
      archivePrefix={arXiv},
      primaryClass={cs.LG},
      url={https://arxiv.org/abs/1801.01290}, 
}

@misc{fujimoto2018addressingfunctionapproximationerror,
      title={Addressing Function Approximation Error in Actor-Critic Methods}, 
      author={Scott Fujimoto and Herke van Hoof and David Meger},
      year={2018},
      eprint={1802.09477},
      archivePrefix={arXiv},
      primaryClass={cs.AI},
      url={https://arxiv.org/abs/1802.09477}, 
}

@article{https://doi.org/10.1002/sam.11583,
author = {Joseph, V. Roshan},
title = {Optimal ratio for data splitting},
journal = {Statistical Analysis and Data Mining: The ASA Data Science Journal},
volume = {15},
number = {4},
pages = {531-538},
doi = {https://doi.org/10.1002/sam.11583},
url = {https://onlinelibrary.wiley.com/doi/abs/10.1002/sam.11583},
eprint = {https://onlinelibrary.wiley.com/doi/pdf/10.1002/sam.11583},
year = {2022}
}

@article{Chen2025PromptLLM,
  author  = {Chen, Qizhao},
  title   = {Explore the Use of Prompt-Based {LLM} for Credit Risk Classification},
  journal = {Journal of Computer and Communications},
  volume  = {13},
  pages   = {33--46},
  year    = {2025},
  doi     = {10.4236/jcc.2025.136003}
}

\end{document}